\documentclass[10pt,twocolumn,letterpaper]{article}

\usepackage[pagenumbers]{cvpr} 

\definecolor{cvprblue}{rgb}{0.21,0.49,0.74}
\usepackage[pagebackref,breaklinks,colorlinks,allcolors=cvprblue]{hyperref}

\usepackage{booktabs}
\usepackage{siunitx}
\usepackage{pifont} 
\usepackage{soul}
\usepackage{multirow}
\usepackage{microtype}

\definecolor{teal}{rgb}{0, 0.5, 0.4}

\definecolor{bestbg}{RGB}{254, 235, 226}   
\definecolor{secondbg}{RGB}{239, 243, 255} 

\hypersetup{urlcolor=pink}
\definecolor{pink}{HTML}{EC008C}

\hypersetup{urlcolor=pink}

\newcommand\blfootnote[1]{
    \begingroup
    \renewcommand\thefootnote{}\footnote{#1} 
    \addtocounter{footnote}{-1}
    \endgroup
}
\def\paperID{2258} 
\def\confName{CVPR}
\def\confYear{2025}

\title{ObjectMover: Generative Object Movement with Video Prior}

\author{
    Xin Yu$^{1,*}$\quad
    Tianyu Wang$^{2}$\quad
    Soo Ye Kim$^{2}$\quad
    Paul Guerrero$^{2}$\quad
    Xi Chen$^{1}$\quad
    Qing Liu$^{2}$\quad
    \\
    Zhe Lin$^{2}$\quad
    Xiaojuan Qi$^{1,\dagger}$\\
    $^{1}$The University of Hong Kong \quad
    $^{2}$Adobe Research \\
    \url{https://xinyu-andy.github.io/ObjMover}
}

\begin{document}
\twocolumn[{%
\renewcommand\twocolumn[1][]{#1}%
\maketitle
\begin{center}
    \centering
    \captionsetup{type=figure}
    \includegraphics[width=0.95\textwidth]{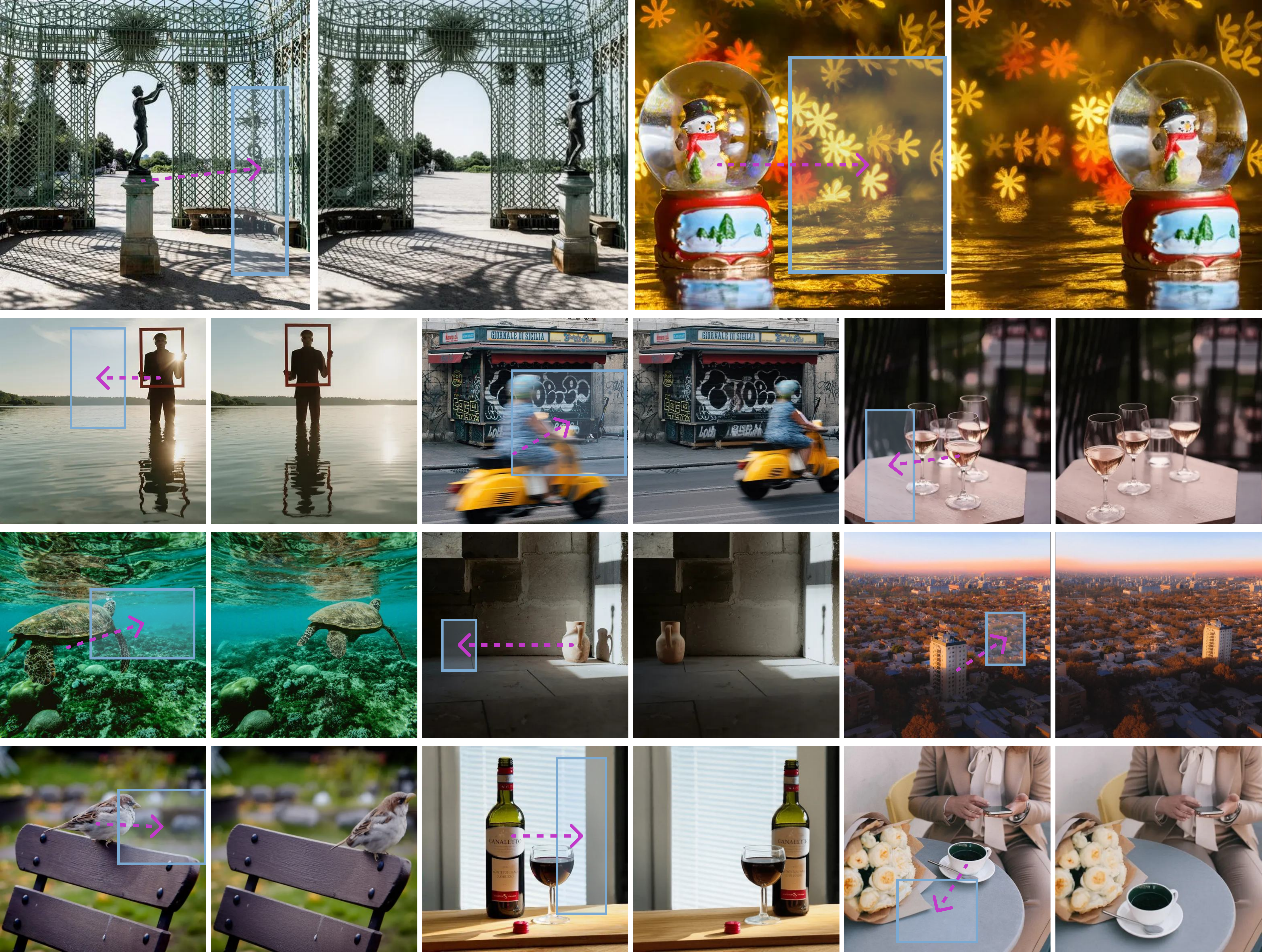}
    \captionof{figure}{\textbf{Results of object movement.} We demonstrate the object movement capability of ObjectMover in a variety of complex and challenging scenarios. ObjectMover can well keep object identity, synchronously edit lighting and shadow effects, complete occluded parts, understand materials, adjust object perspective, and comprehend occlusion relationships to generate realistic images where the object has been moved, maintaining all other elements of the scene unchanged. } 
    \label{fig:teaser}
\end{center}%
}]

\blfootnote{* Work done during an internship at Adobe.}
\blfootnote{$\dagger$ Corresponding authors.}

\begin{abstract}
Simple as it seems, moving an object to another location within an image is, in fact, a challenging image-editing task that requires re-harmonizing the lighting, adjusting the pose based on perspective, accurately filling occluded regions, and ensuring coherent synchronization of shadows and reflections
while maintaining the object identity. 
In this paper, we present \textit{ObjectMover}, a generative model that can perform object movement in highly challenging scenes. Our key insight is that we model this task as a sequence-to-sequence problem and fine-tune a video generation model to leverage its knowledge of consistent object generation across video frames.
We show that with this approach, our model is able to adjust to complex real-world scenarios, handling extreme lighting harmonization and object effect movement.
As large-scale data for object movement are unavailable, we construct a data generation pipeline using a modern game engine to synthesize high-quality data pairs.
We further propose a multi-task learning strategy that enables training on real-world video data to improve the model generalization. Through extensive experiments, we demonstrate that \textit{ObjectMover} achieves outstanding results and adapts well to real-world scenarios.

\end{abstract}    

\section{Introduction}
\label{sec:intro}

Moving an object within an image is a fundamental task in the field of image editing, with applications spanning photography enhancement, content creation, and entertainment. This seemingly simple task is inherently complex as it needs to preserve the object's identity, seamlessly fill occluded regions, and update effects like shadows, reflections, and lighting to be consistent to the new object location (see \cref{fig:teaser}).
Additionally, to realistically render the object movement in a 3D scene as a 2D image, the perspective view of the object often needs to change or more generally, the pose of the object and its new surroundings may need to adapt to each other (see the last three examples in \cref{fig:teaser}).

\begin{figure}
    \centering
    \includegraphics[width=0.98\linewidth]{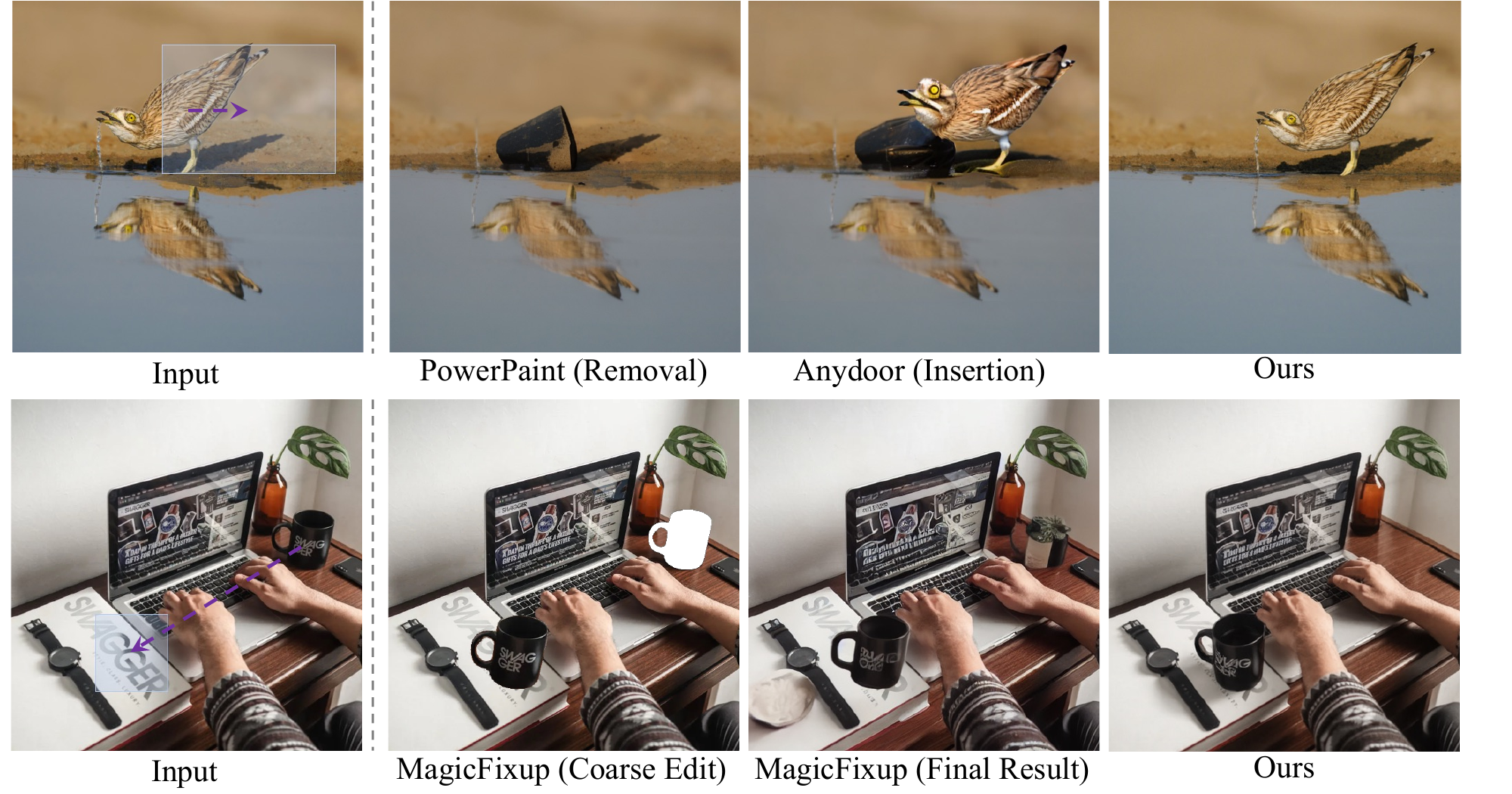}
    \vspace{-0.1in}
    \caption{\textbf{Limitations of existing methods for object movement.} (Top): The approach of removing an object~\cite{powerpaint} and then re-inserting~\cite{anydoor} it for repositioning leads to issues with identity preservation and ineffective synchronization of lighting effects. (Bottom): The copy-paste-based method~\cite{magic_fixup} is unable to modify the perspective of the object.}
    \vspace{-0.2in}
    \label{fig:intro}
\end{figure}

Despite significant advancements in image editing~\cite{anydoor,objectdrop,objectstitch,imprint,powerpaint,paint_by_example,3dit,li2022image,mat}, few methods can directly accomplish the task of object movement. A natural approach to solve this problem is to formulate it as two sequential image editing tasks: object removal~\cite{powerpaint,lama} at the source location and object insertion~\cite{objectstitch,imprint,anydoor,paint_by_example} at the target location. However, for a complete removal of the object at its source location, first defining an appropriate region covering the object with its associated effects such as shadows and reflections can already be highly challenging. Moreover, as object insertion models are not explicitly trained for object movement within the same image, they will often noticeably modify the identity of the object and have inconsistent lighting or shadows compared to the object in the original location (see top row in \cref{fig:intro} for artifacts resulting from this two-step approach).
Another two-step approach involves first manually copy-and-pasting the object to a new location, followed by a model to harmonize it with the surrounding environment~\cite{objectdrop,magic_fixup}. However, this fails to account for natural perspective changes for the object at the target location due to the naive copy-and-pasting and often fails to harmonize shadows (see \cref{fig:intro}, bottom).

\begin{figure}
    \centering
    \includegraphics[width=0.95\linewidth]{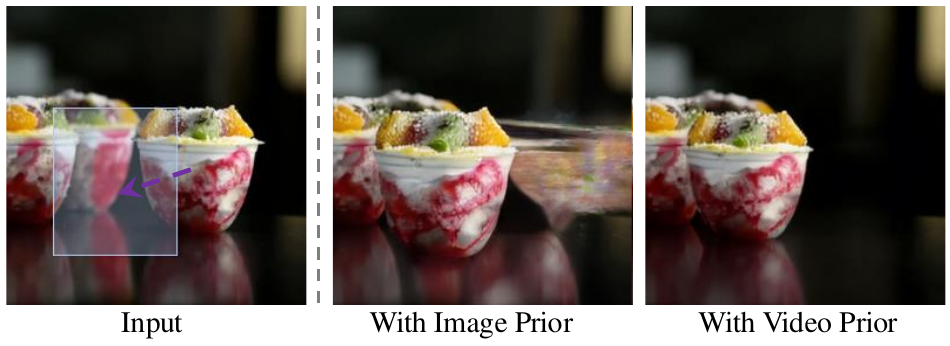}
    \vspace{-0.1in}
    \caption{\textbf{Results trained with different priors.} The model fine-tuned from a video prior outperforms the one from an image prior.}
    \vspace{-0.2in}
    \label{fig:intro_ablation}
\end{figure}

To address these limitations, we present \textit{ObjectMover}, a single-stage approach for object movement based on a diffusion transformer model.
Unlike conventional methods that rely on pre-trained text-to-image models for image editing~\cite{anydoor,objectstitch,objectdrop,imprint,magic_fixup}, our approach reformulates the task as a sequence-to-sequence prediction problem and repurposes a pre-trained video diffusion model to solve the problem. By treating the input scene image, object of interest, user instructions, and target frame as a sequence of frames, our method can leverage learned video priors to capture the consistent evolution of lighting, object identity, and scene context across frames. This enables natural, consistent results, which image generative models -- lacking such priors -- often fail to achieve (see \cref{fig:intro_ablation}).

One main challenge in training is the lack of paired data. To tackle this, we propose to use two complementary sources of data for training.
Firstly, we propose to utilize game engines like Unreal Engine~\cite{UE} to generate high-quality, aligned synthetic training data specifically designed for movement, which is crucial for transferring and adapting video prior to the target task. 
Another source of data is from incorporating natural videos to enhance the generalization and improve the performance in real-world scenarios. 
Despite not being perfectly aligned with our task due to issues like changing backgrounds and hence not directly usable for object movement, we adopt a multi-task learning strategy and train with videos on an auxiliary task of mask-based insertion, which enriches the model with real-world domain knowledge, leading to high-quality results, strong generalization ability (see \cref{tab:ablation} and \cref{fig:ablation}).
Meanwhile, we fully utilize the synthetic data by training on mask-free object insertion and removal, which equips our model with the ability to address multiple tasks within a single framework (see \cref{fig:removal} and \cref{fig:insertion}).

Through extensive experiments, we demonstrate that \textit{ObjectMover} achieves state-of-the-art results and outperforms existing methods by a large margin (see \cref{fig:comparison} and \cref{tab:comparison}).
In summary, our contributions are:

\begin{itemize}
    \item We introduce a novel framework that models object movement as a sequence-to-sequence problem and repurpose a video model for single-image editing tasks.
    
    \item To address data scarcity, we develop a synthetic data construction pipeline using modern game engines which produces a high-quality, high-resolution dataset suitable for object movement, removal, and insertion.
    
    \item We propose a multi-task learning pipeline that fully leverages different data sources across relevant tasks, enhancing target task performance with real-world video data and equipping the model with versatile capabilities in object removal and insertion.
    

    \item Through extensive experiments, we demonstrate that \textit{ObjectMover} achieves superior results and outperforms existing methods.
\end{itemize}

\section{Related work}
\label{sec:related_work}

Most related to our approach are image editing methods for generative removal, insertion, and movement.
Additionally, motion-controlled video generators could implement generative movement by generating a video of the moving object.

\vspace{0.05in}\noindent{\textbf{Generative Removal \& Insertion.}} Inpainting methods based on diffusion models~\cite{avrahami2022blended, suvorov2021resolution, lugmayr2022repaint, meng2021sdedit, saharia2022palette} are currently state-of-the-art for generative removal, where the object is removed by inpainting image region covered by the object.
Similarly, generative insertion~\cite{lu2023tficon, zhang2023controlcom, anydoor, paint_by_example, objectstitch, imprint} is currently dominated by diffusion-based inpainting that additionally guide the generated content to match a given object through identity preservation approaches.
Most of these techniques maintain identity only roughly, which is particularly problematic for generative movement, where even minor discrepancies are noticeable.
Furthermore, all inpainting-based insertion and removal methods fail to create or eliminate effects outside the inpainting mask, such as long shadows, reflections, or indirect lighting. Expanding the inpainting mask results in the loss of the original image's identity, as it is not maintained within the inpainting mask.
To tackle this, certain methods focus on removing only shadows while preserving other scene attributes~\cite{guo2023shadowdiffusion, wang2020instance}, but they cannot eliminate other effects like reflections or indirect lighting, and they rely on an accurate shadow mask, which is challenging to obtain reliably.
Recently, ObjectDrop~\cite{objectdrop} has been proposed as a mask-free object insertion and removal method, allowing one to address object movement in a two-step manner (i.e., first remove and then re-insertion). Nevertheless, this method primarily involves copying and pasting, treating movement as two separate editing tasks. This approach may lead to inconsistencies, unnatural perspectives, poses, or changes in lighting. Another recent insertion method~\cite{think_outside} allows changes across the whole image when inserting objects, yet it does not support removal or movement.

\vspace{0.05in}\noindent{\textbf{Generative Movement.}} Several image editing methods have recently been proposed to move an object shown in an image in a semantically plausible way. Typically, they generate an edited image by adding identity preservation and some form of control mechanism to a diffusion model. A popular control mechanism is a drag-based interface~\cite{mou2023dragondiffusion, shi2024dragdiffusion, avrahami2024diffuhaul, mou2024diffeditor}, where the user defines where specific points in the image should end up. Other forms of control include motion fields~\cite{geng2024motion}, a coarse version of the edit~\cite{magic_fixup, objectdrop}, or 3D-aware controls~\cite{3dit, diff_handle, yenphraphai2024image, bhat2023loosecontrol} that also allow for user-driven changes to the 3D pose and perspective of an object.
However, none of the methods have demonstrated strong context-driven changes of the object being moved, such as changes in perspective or lighting.

\vspace{0.05in}\noindent{\textbf{Motion-Controlled Video Generators.}} Motion control for video generation shares the goal of moving an object in a scene. Recent methods allow for specifying the motion directions of full trajectories of objects in an image~\cite{shi2024motion, wang24boximator}. However, trained on video data, these methods do not have a mechanism to ensure that unrelated parts of the scene do not change, such as lighting or secondary objects. The recent DragAnything~\cite{drag_anything} addresses this issue to some extent, but is still hard to achieve background preservation and context-driven changes compared to our approach (see \cref{fig:comparison}).

\section{Method}
\label{sec:method}

\begin{figure*}
    \centering
    \includegraphics[width=1\textwidth]{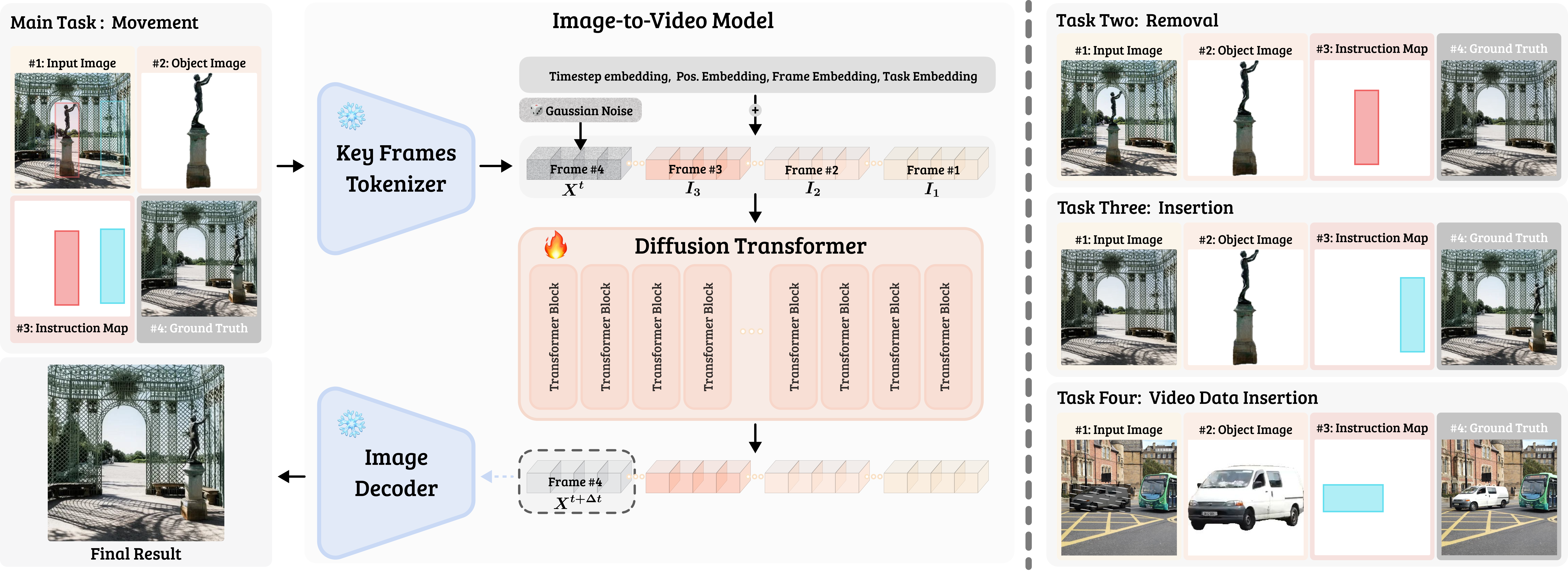}
    \vspace{-0.1in}
    \caption{\textbf{Model architecture.} (Left): Our overall sequence-to-sequence framework by leveraging an image-to-video prior for training our single-frame image editing task (\cref{sec:3.1}). (Right): Frame formulation on different tasks for multi-task learning (\cref{sec:3.3}).}
    \vspace{-0.2in}
    \label{fig:model}
\end{figure*}

\subsection{Overview}
There are two main goals for the object movement problem: (i) the complete removal of the object at the source location including its associated effects such as shadows or reflections, and (ii) the consistent insertion of the object at the target location. More specifically, the generated object should retain an identity that aligns with the source object while ensuring that its interactions with the surrounding environment are accurately updated. This often involves the appearance of a new shadow in the target area and its disappearance from the original location due to movement. For instance, when an object moves from sunlight to shade, it should blend with the darker environment; similarly, a person's shadow should be cast on the wall when they are closer to it. It is important to note that when there are several plausible ways of harmonizing an object at a new location to make the image look natural, unlike with object insertion, the harmonization in object movement should match the object at the original location to maintain consistent physical interaction between the static scene and moving objects.

Our key insight is that object movement can be considered a special case of a two-frame video, where the object consistency and dynamic effect evolution between different frames have already been inherently learned in video generation models, as it was trained by watching multi-frame real-world events. 
Thus, instead of following conventional image-editing approaches that rely on an image diffusion model, we propose to leverage a video model as our foundation. Our approach consists of three primary components: First, we reformulate the task of object movement as a sequence-to-sequence prediction problem, enabling us to repurpose and fine-tune a pre-trained image-to-video model (\cref{sec:3.1}). Second, to fine-tune the pre-trained video diffusion model for the target task, we use a game engine to construct a task-aligned synthetic dataset that is sufficiently diverse and capable of simulating instructive data pairs not typically found in video datasets (\cref{sec:3.2}). Lastly, to fully leverage the potential of synthetic data and also real-world video data, we employ multi-task learning, incorporating related tasks that can be trained on both types of data in a unified manner, to enhance the generalization capability of our model (\cref{sec:3.3}).

\subsection{Repurposing Video Diffusion Models}
\label{sec:3.1}

We utilize a pre-trained image-to-video model based on a diffusion transformer architecture~\cite{transformer,dit} similar to Sora~\cite{sora} and devise a sequence-to-sequence framework that is consistent with the original video model input, even though we only target at a single-frame image generation. As depicted in \cref{fig:model}, we formulate all conditional information, including editing instructions, as image frames: (1) an input image $I_1$; (2) a foreground object image $I_2$; and (3) an instruction map $I_3$. The instruction map $I_3$ is formed by layering two bounding boxes across separate channels such that $I_3 = [M_{src}, M_{tar}, M_{tar}]$ with $M_{src}$ signifying the initial position and $M_{tar}$ the target position. Since we use a latent-diffusion~\cite{latent_diffusion} model, all image frames are converted into tokens within the latent space via a VAE encoder. Notationally, unless specified otherwise, the symbols $I_1$, $I_2$, $I_3$, and related terms will refer to the VAE-encoded tokens rather than the original images.

During inference, starting from pure noise $X^0 \sim \mathcal{N}(0,1)$, a noise frame $X^t$ is introduced to combine with condition images to create a sequence $S^t=[I_1,I_2,I_3,X^t]$ and our model $v_{\theta}$ iteratively denoises this sequence and finally predicts the clean image $X^1$. At each denoising step, the model also produces an output sequence of equal length, $S^{t+\Delta t}:=[I^{t+\Delta t}_1, I^{t+\Delta t}_2, I^{t+\Delta t}_3, X^{t+\Delta t}]$. Before each subsequent denoising step, $I^{t+\Delta t}_1, I^{t+\Delta t}_2, I^{t+\Delta t}_3$ are replaced back with $I_1, I_2, I_3$.

The training employs a flow-matching loss~\cite{flow_matching,instaflow}. Specifically, we use linear interpolation to infuse noise into the ground-truth target, generating the noisy input as follows:
\begin{equation}
    X^t = tX^1 + (1 - t)X^0,
\end{equation}
where $X^1$ is the ground-truth image and $X^0 \sim \mathcal{N}(0,1)$ represents Gaussian noise. The model is designed to predict the velocity $V^t = \frac{dX^t}{dt} = X^t - X^0$, which can be converted to an estimated $X^{t+\Delta t}$ during inference. The loss function is formulated as:
\begin{equation}
\mathcal{L}=\mathbb{E}_{t, X^0, X^1}\left\|v\left(S^t, t ; \theta\right)_{[4]}-V^t\right\|^2,
\end{equation}
where $v_{[4]}$ is the fourth frame of the output sequence, focusing the loss calculation on the target frame.

A key challenge, however, is the lack of access to the ground-truth image $X^1$. In \cref{sec:3.2}, we present a synthetic data simulation pipeline to support training, and in \cref{sec:3.3}, we explore multi-task training on unpaired video data.

\subsection{Data Generation with a Game Engine}
\label{sec:3.2}
\begin{figure}
    \centering
    \includegraphics[width=1\linewidth]{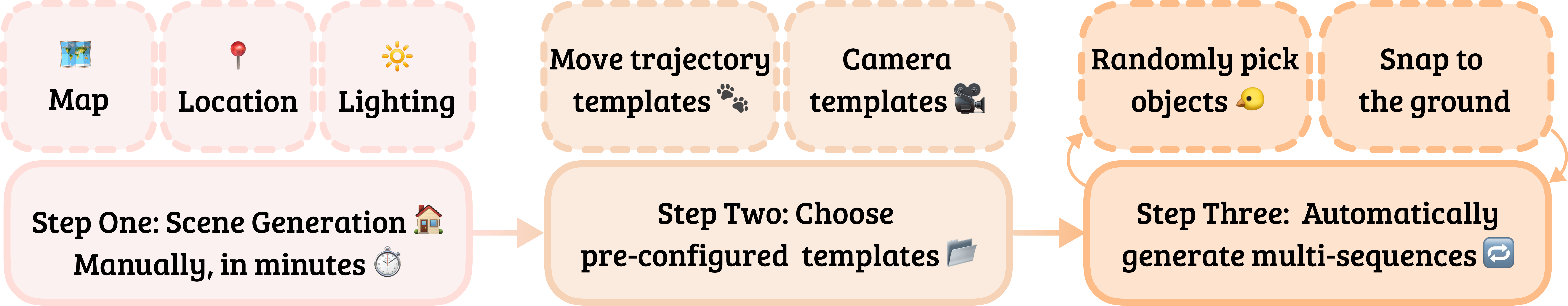}
    \vspace{-0.1in}
    \caption{\textbf{Data generation pipeline.} The pipeline consists of three steps: (1) background scene generation; (2) movement template pre-configuration; and (3) object movement sequence generation.}
    \vspace{-0.2in}
    \label{fig:data_tool}
\end{figure}

As large-scale paired data for object movement are unavailable, we propose to construct the training data with a game engine, Unreal Engine~\cite{UE}. As shown in \cref{fig:data_tool}, our data generation pipeline consists of three steps: (1) background scene generation; 
(2) movement template pre-configuration;
and (3) object movement sequence generation.

\begin{figure}
    \centering
    \includegraphics[width=1\linewidth]{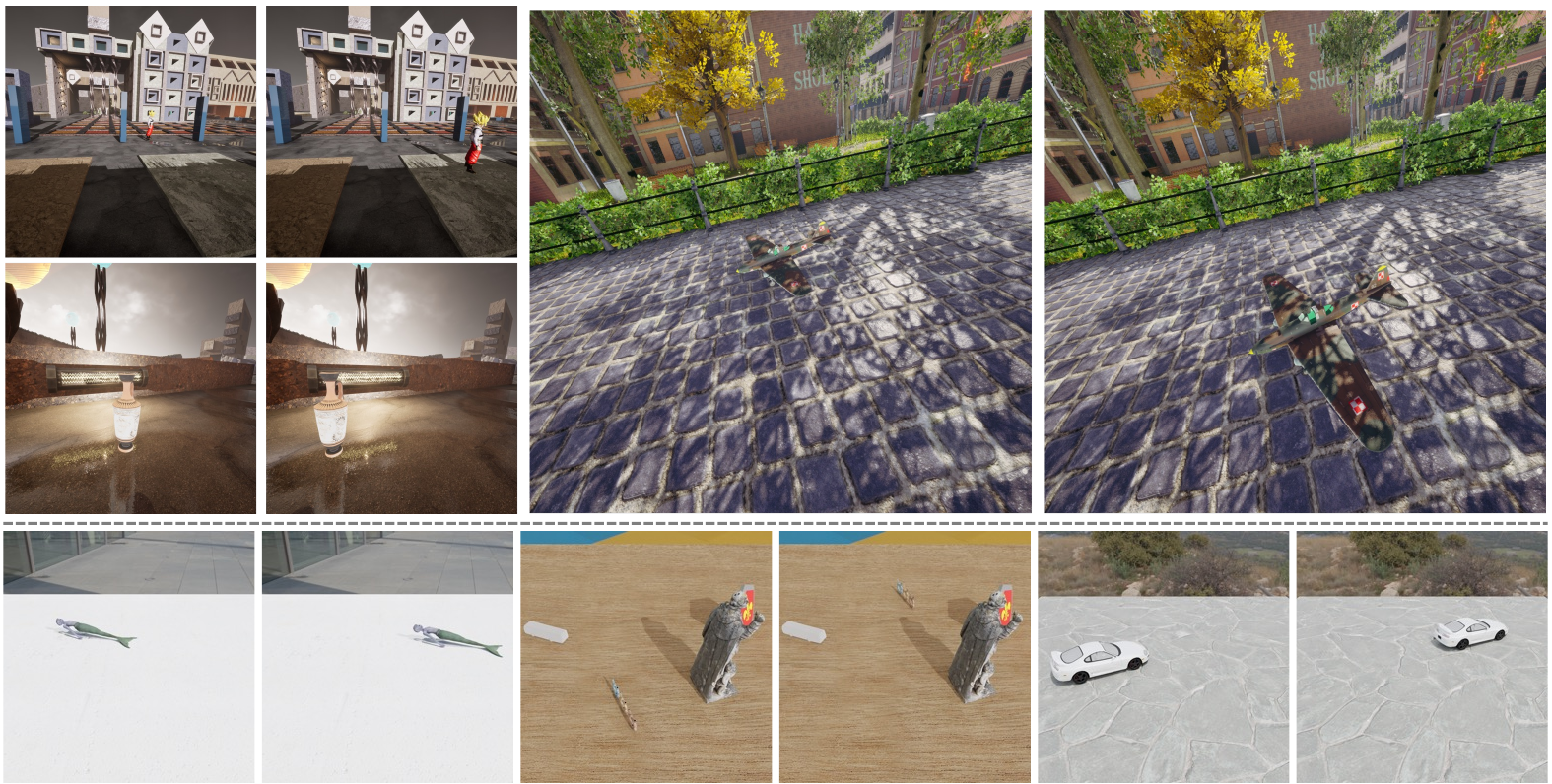}
    \vspace{-0.2in}
    \caption{\textbf{Dataset comparison.} Row (1-2) shows our synthetic data using a game engine; Row (3) shows existing synthetic data~\cite{3dit}. Our data is more realistic and has complex lighting effects.}
    \vspace{-0.1in}
    \label{fig:data}
\end{figure}

\vspace{0.05in}\noindent{\textbf{Step 1: Background Scene Generation.}} For the background scene, we use a collection of maps intended for gaming purposes to enhance realism, different from earlier simpler configurations~\cite{3dit} that depended on manually constructed scenes featuring a limited number of scene objects and simplistic environmental maps, as shown in \cref{fig:data}. Using these maps has the following advantages: (1) scene particulars including lighting, background meshes, and assets are meticulously orchestrated by specialists to ensure high fidelity, resulting in realistic shadows, reflections, and background scenes; (2) expansive maps that allow versatile object placement for rendering across various locations within the same environment where diverse background images can be created even from a single map; (3) adjustable lighting parameters, such as intensity, temperature, and direction, for diverse lighting conditions.

\vspace{0.05in}\noindent{\textbf{Step 2: Movement Template Pre-configuration.}}
Manually placing objects at multiple locations within a background scene can be extremely time-consuming. To address this, we develop a variety of trajectory and camera position templates to automate object placement in $k$ intermediate locations along the trajectory, capturing intermediate frames in the process. Notably, direct interpolation and placement of objects can lead to clipping artifacts when objects penetrate the ground, violating physical placement principles. To prevent this, after determining each location, we perform ray casting to detect any intersections between the object and the ground (or other objects). We then automatically adjust the position and orientation to ensure the object’s bottom surface aligns accurately with the ground. This adjustment is dynamically influenced by the map's local terrain, providing more variation and enhancing data diversity. This step is a fully automated process to facilitate large-scale data generation.

\vspace{0.05in}\noindent{\textbf{Step 3: Object Movement Sequence Generation.}} Finally, with different scene maps, objects and movement trajectories, we generate 18,783 distinct sequences with frame lengths of $k=10$ or $k=15$, including one frame with a clean background devoid of objects for mask-free object insertion and removal training (see \cref{sec:3.2}). This results in a total of 1,348,248 training pairs, $13\times$ greater than the previous work~\cite{3dit}. Furthermore, the images are captured in high resolutions ($1024\times1024$ and $1080\times1920$), unlike earlier datasets~\cite{3dit} ($256\times256$), and thus supporting our high-resolution training for a more practical usage.

\begin{figure*}[t]
    \centering
    \includegraphics[width=0.98\textwidth]{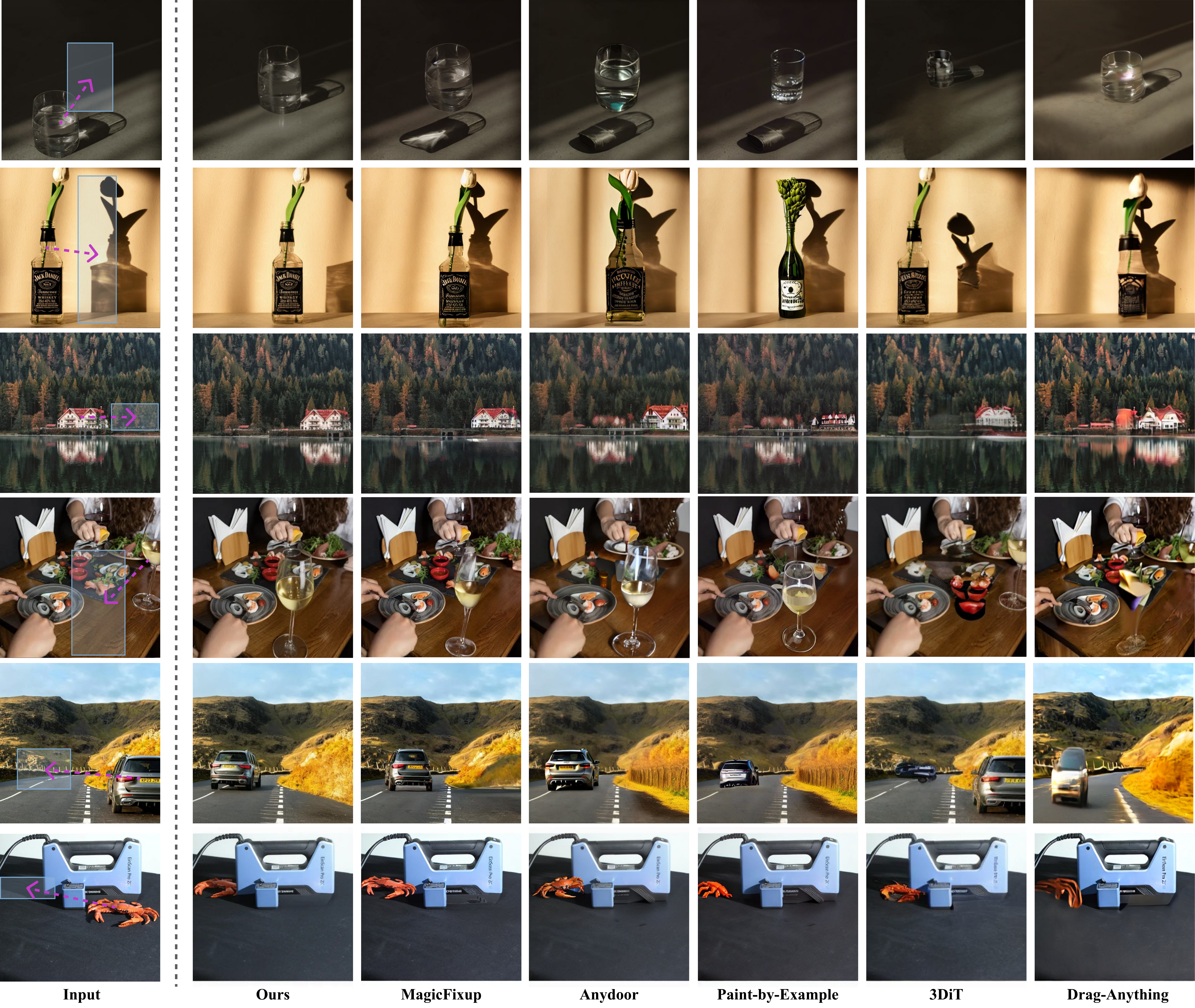}
    \vspace{-0.1in}
    \caption{\textbf{Qualitative comparisons on object movement.} Our method consistently outperforms state-of-the-art methods in maintaining object identity, lighting consistency, and overall quality. Notably, in the last example, only our method successfully harmonizes with the target region, achieving a realistic effect with partial shadows and light.}
    \vspace{-0.2in}
    \label{fig:comparison}
\end{figure*}

\subsection{Training with Video via Multi-Task Learning}
\label{sec:3.3}

The benefit of the synthetic data generation pipeline described in Section~\ref{sec:3.2} is clear: it enables the collection of diverse paired data dedicated to object movement under various lighting scenarios through intentional control of lighting parameters. Nevertheless, a limitation of the synthetic data is that it may still exhibit a domain gap from the natural image distribution and lack the real-world diversity and complexities of dynamic changes across frames. Hence, a good source of data for object movement can be natural videos with moving objects. For video data to serve our task, ideally, everything would remain static except for the moving object; however, this is not the case in most videos. Thus, we devise a multi-task learning (MTL) strategy that allows us to use both real video data and synthetic data.

\vspace{0.05in}\noindent{\textbf{MTL on Synthetic Data.}} Our synthetic data includes full background images without the object, which not only supports object movement but also facilitates object removal and insertion tasks. It should be noted that existing compositing methods \cite{objectstitch, anydoor} require masking out the background image, leading to significant limitations: a restricted generation scope, where the model can only generate within the masked area, thus preventing the creation of more realistic effects such as shadows and reflections; and unintended background alterations, where regenerating within the mask inadvertently alters original background regions. In contrast, our approach does not require pre-masking the target location because full background images are available in our synthetic dataset, thereby avoiding the aforementioned limitations. On the synthetic data, our model is jointly trained for all three tasks of object removal, insertion, and movement.

\vspace{0.05in}\noindent{\textbf{MTL on Video Data.}}
As we lack full background frames for the video data, we incorporate an auxiliary task of \textit{mask-based} insertion (i.e., masking out the background image as in conventional compositing models, wherein we replace the pixels within the mask region with black pixels) to integrate the video data into our training regimen. This enables our model to adapt to real-world content by training in natural videos and helps improve synthesize complex lighting effects. As illustrated on the right side of \cref{fig:model}, our sequence-to-sequence framework consolidates the training of various tasks. In this framework, the edited image consistently corresponds to $I_1$, the segmented object to $I_2$, and the instruction map to $I_3$. Specifically, for removal tasks, only the source location mask is employed in $I_3$, while for insertion tasks, only the target location mask is utilized. 
For the insertion task, we use the object image from a frame different from the target image to compel the model to adapt to the new environment. 
Across all tasks, the noisy target input is positioned in the final frame. Besides, a unique task embedding is developed for each distinct task.

\section{Experiments} 
\label{sec:experiment}

\begin{figure}[t]
    \centering
    \includegraphics[width=1\linewidth]{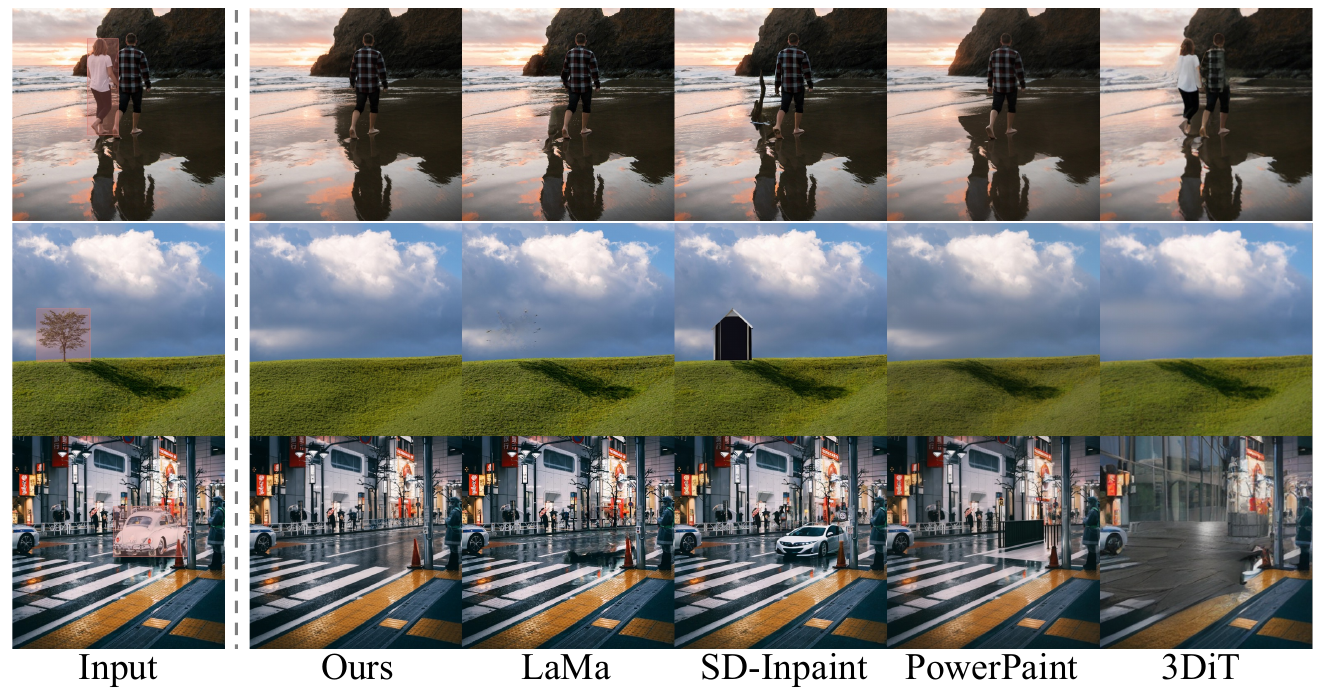}
    \vspace{-5mm}
    \caption{\textbf{Qualitative comparisons on object removal.} Our method consistently outperforms state-of-the-art methods.}
    \vspace{-0.2in}
    \label{fig:removal}
\end{figure}

\vspace{0.05in}\noindent{\textbf{Training Details.}} We train ObjectMover using an image-to-video model that is based on a transformer architecture and has 5B parameters. Note that the method can be applied to any video diffusion model. We employ the AdamW~\cite{adam} optimizer with a weight decay of 0.01 and set the initial learning rate to $1 \times 10^{-4}$. The model is trained at a resolution of $512 \times 512$ pixels with a batch size of 256. EMA with a decay rate of 0.9999 is applied starting after the first 1,000 iterations to stabilize training. Training involves both synthetic and video datasets, with a 1:1 ratio. Tasks in the synthetic dataset are divided into object movement, removal, and insertion at a 6:1:1 ratio. We leave the data processing of video data in the supplementary file. Box augmentation is employed by maintaining the bounding box's center and adjusting its dimensions randomly within set ranges (i.e., 0.8-1.2 on synthetic data; 1.0-1.5 on video data for full mask-out), which improves the model’s robustness.

\vspace{0.05in}\noindent{\textbf{Evaluation Details.}} For evaluation, we introduce two new datasets: \textit{ObjMove-A} and \textit{ObjMove-B}. \textit{ObjMove-A} comprises 200 image sets captured by experienced photographers. Each set includes an object placed at two distinct locations within the same scene, alongside a pure background image without the object. Object masks are subsequently generated using SAM~\cite{sam}. Thus, the model performance can be quantitatively assessed between the generated image and the referenced target image on this dataset. Following~\cite{anydoor,think_outside,imprint}, we utilize object-level metrics such as DINO-Score~\cite{dino}, CLIP-Score~\cite{clip}, and DreamSim~\cite{dreamsim} to evaluate the cropped target region. Since our task involves removing the object from the source region, we also evaluate these metrics on the cropped source region, and we show the average score on these two regions. We also assess the PSNR for the full generated image to measure overall image similarity.

Given that our task is inherently ill-posed and reference-based evaluations cannot fully capture the realism of the generated images, we conduct human evaluations using our \textit{ObjMove-B} dataset. This dataset consists of 150 image sets sourced from Pexels~\cite{Pexels}, consisting of manually annotated object masks and bounding box masks for plausible target regions. Compared with \textit{ObjMove-A}, this dataset presents more diverse scenes and intricate challenges, such as occlusions, complex shadows and reflections, and complex background, thereby providing a more challenging benchmark for assessment. To evaluate the result, we conduct a user study in which participants select the best result from different methods based on identity preservation, consistent lighting effects editing, and overall image quality.

\begin{table}[t]
\centering
\resizebox{\linewidth}{!}{
\begin{tabular}{l c c c c c}
\toprule
\textbf{Method} & \textbf{PSNR} $\uparrow$ & \textbf{DINO-Score} $\uparrow$ & \textbf{CLIP-Score} $\uparrow$ & \textbf{DreamSim} $\downarrow$ & \textbf{User-Study} $\uparrow$\\
\midrule
Drag-Anything~\cite{drag_anything} & 16.36 & 55.56 & 84.44 & 0.411 & 0.19\%\\
3DiT~\cite{3dit} & 19.72 & 45.30 & 81.69 & 0.514 & 0.19\%\\
Paint-by-Example~\cite{paint_by_example} & 20.83 & 55.46 & 85.23 & 0.420 & 0.75\%\\
Anydoor~\cite{anydoor} & 21.86 & 69.32 & 88.95 & 0.289 & 3.56\%\\
MagicFixup~\cite{magic_fixup} & \underline{23.82} & \underline{78.49} & \underline{91.06} & \underline{0.198} & \underline{31.14\%}\\
Ours & \textbf{25.27} & \textbf{85.07} & \textbf{93.16} & \textbf{0.142} & \textbf{64.17\%}\\
\bottomrule
\end{tabular}
}
\caption{\textbf{Quantitative comparisons on object movement.} Our method consistently outperforms state-of-the-art methods. For the evaluation metrics, only the user-study is conducted on \textit{ObjMove-B}; all other evaluations are performed using \textit{ObjMove-A}, as only \textit{ObjMove-A} has ground-truth data.}
\vspace{-0.1in}
\label{tab:comparison}
\end{table}

\subsection{Comparison to Existing Methods}

\vspace{0.05in}\noindent{\textbf{Comparison on Object Movement.}} We first evaluate our model on the core task of object movement against the most relevant approaches~\cite{magic_fixup, 3dit, drag_anything, anydoor, paint_by_example}. Specifically, Anydoor~\cite{anydoor} and Paint-by-Example~\cite{paint_by_example} are limited to object insertion and cannot relocate existing objects. Therefore, we employ the state-of-the-art inpainting model PowerPaint~\cite{powerpaint} to remove the object from its original position and subsequently utilize these methods to insert the object into the target location. For Drag-Anything~\cite{drag_anything}, we generate a video following a linear trajectory from the source location to the target location and save the last frame as the edited result (see Supplementary for implementation details).

\begin{figure}
    \centering
    \includegraphics[width=1\linewidth]{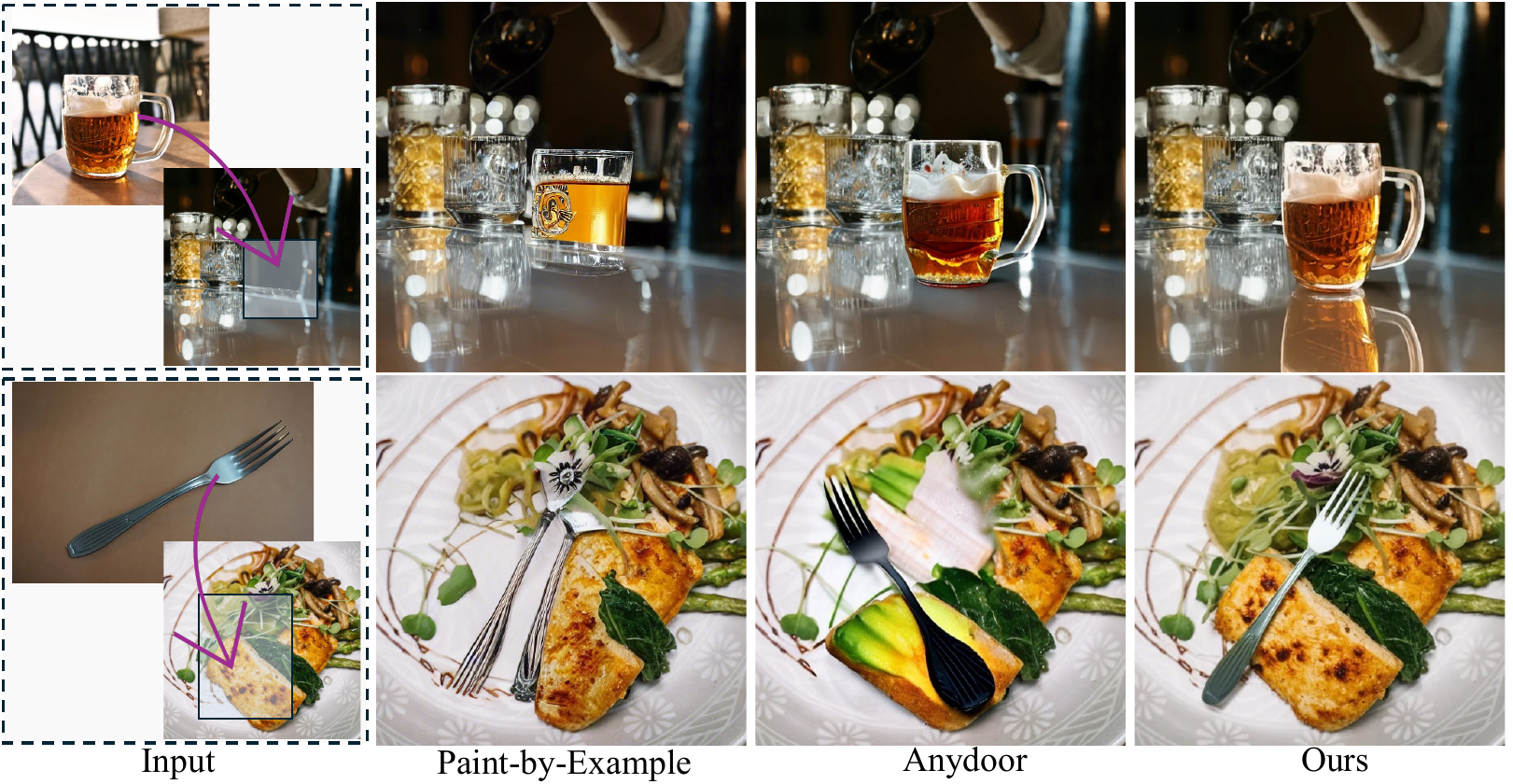}
    \vspace{-0.25in}
    \caption{\textbf{Qualitative comparisons on object insertion.} Our method consistently outperforms state-of-the-art methods.}
    \vspace{-0.2in}
    \label{fig:insertion}
\end{figure}

\cref{tab:comparison} shows a quantitative analysis, where our approach consistently surpasses existing methods. Qualitative comparisons in \cref{fig:comparison} highlight our model's significant superiority. Specifically, in the first three examples, competing methods fail to effectively adjust object shadows or reflections to the target location. Furthermore, Anydoor and Paint-by-Example struggle with preserving object identity. In the challenging fourth example, involving completing an incomplete object and understanding glass material transparency requiring background synchronization, our method excels. Other methods either leave the object incomplete or distort the background. For the fifth image, where changing the car's perspective angle to suit the new location is crucial, MagicFixup fails, and Anydoor and Paint-by-Example alter the object's identity. In the final example, only our method successfully integrates into the target region using lighting information, achieving a realistic effect of partial shadow and light. We also perform a self-evaluation game (details in the supplementary file), presenting four images: one is the input image, and the other three are generated by our model. Users are asked to identify the input image and the results reveal that users incorrectly identify the input image 70\% of the time, demonstrating our method's ability to generate images that obscure artifacts.


\vspace{0.05in}\noindent{\textbf{Comparison on Auxiliary Tasks.}}
Our model is capable of object removal and insertion, and we compare these capabilities with other methods~\cite{lama,latent_diffusion,powerpaint,3dit,anydoor,paint_by_example}. For object insertion, we use the segmented object from the source frame as a reference image and prompt the model to place the object into the clean background at the target frame's location. We evaluate object-centric metrics in the cropped target region for insertion and in the source region for removal. \cref{tab:auxiliary} shows quantitative comparisons, while \cref{fig:removal} and \cref{fig:insertion} provide visual comparisons. Our results indicate that our model outperforms other approaches. Specifically, in object removal, our method effectively removes the object without introducing artifacts, eliminating shadows and reflections more successfully than others, and show stronger background completion ability. In object insertion, other methods struggle with consistent reflections, identity, and background preservation, while our approach maintains consistent reflections, preserves identity and background integrity, and blends seamlessly with the environment.

\begin{table}[t]
\centering
\resizebox{\linewidth}{!}{
\small 
\begin{tabular}{l c c c c }
\toprule
\textbf{Method} & \multirow{2}{*}{\textbf{PSNR} $\uparrow$} & \multirow{2}{*}{\textbf{DINO-Score} $\uparrow$} & \multirow{2}{*}{\textbf{CLIP-Score} $\uparrow$} & \multirow{2}{*}{\textbf{DreamSim} $\downarrow$} \\
\textit{Object Removal} &&&& \\
\midrule
3DiT~\cite{3dit} & 20.52 & 41.83 & 85.02 & 0.499  \\
PowerPaint~\cite{powerpaint} & 25.15 & 62.71 & 89.40 & 0.347  \\
SD-Inpaint~\cite{latent_diffusion} & 24.16 & 50.80 & 83.91 & 0.492  \\
LaMa~\cite{lama} & \underline{27.04} & \underline{63.82} & \underline{89.91} & \underline{0.322}  \\
Ours & \textbf{28.90} & \textbf{83.94} & \textbf{94.84} & \textbf{0.143}  \\
\midrule
\textit{Object Insertion} & \textbf{PSNR} $\uparrow$ & \textbf{DINO-Score} $\uparrow$ & \textbf{CLIP-Score} $\uparrow$ & \textbf{DreamSim} $\downarrow$ \\
\midrule
Paint-by-Example~\cite{paint_by_example} & 22.64 & 54.80 & 81.81 & 0.457  \\
Anydoor~\cite{anydoor} & \underline{24.10} & \underline{77.49} & \underline{88.33} & \underline{0.223}  \\
Ours & \textbf{24.99} & \textbf{85.69} & \textbf{91.45} & \textbf{0.152}  \\
\bottomrule
\end{tabular}
}
\vspace{-0.1in}
\caption{\textbf{Quantitative comparisons on auxiliary tasks.} Our method consistently outperforms state-of-the-art methods.}
\vspace{-0.2in}
\label{tab:auxiliary}
\end{table}
\begin{table}[t]
\centering
\resizebox{\linewidth}{!}{
\small 
\begin{tabular}{l c c c c }
\toprule
\textbf{Method} & \textbf{PSNR} $\uparrow$ & \textbf{DINO-Score} $\uparrow$ & \textbf{CLIP-Score} $\uparrow$ & \textbf{DreamSim} $\downarrow$ \\
\midrule
Model A & 22.86 & 75.01 & 92.19 & 0.242 \\
Model B & 23.83 & \underline{80.02} & 93.16 & \underline{0.196} \\
Model C & \underline{23.89} & 79.97 & \underline{93.22} & 0.197 \\
Model D & \textbf{24.24} & \textbf{82.09} & \textbf{93.62} & \textbf{0.180} \\
\bottomrule
\end{tabular}
}
\caption{\textbf{Ablation of different models.} Model A indicates training from the T2I checkpoint with synthetic data only. Model B indicates training from the I2V checkpoint with synthetic data only and only on the movement task. Model C indicates training from the I2V checkpoint with synthetic data only but adding the insertion and removal tasks. Model D indicates adding video data.}
\vspace{-0.2in}
\label{tab:ablation}
\end{table}

\subsection{Ablation Studies}

\begin{figure}
    \centering
    \includegraphics[width=0.95\linewidth]{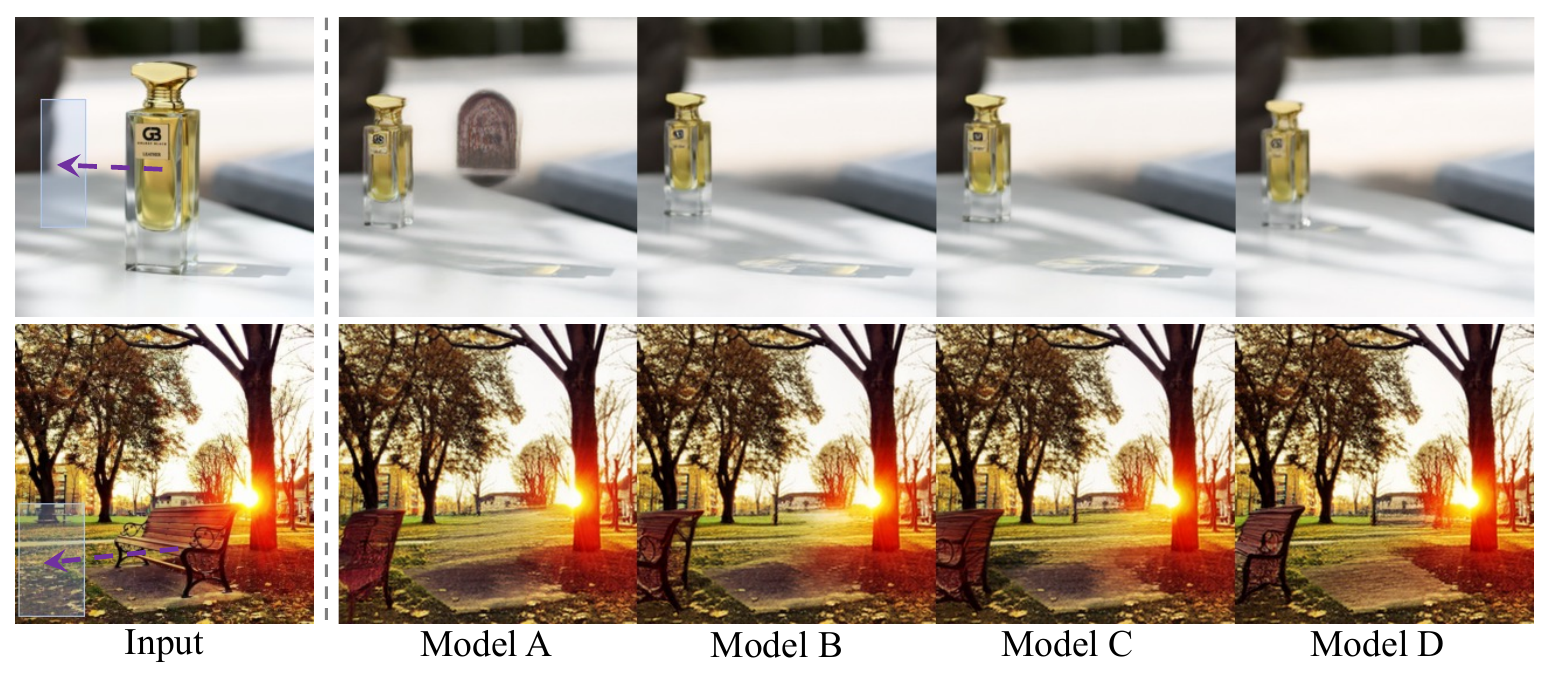}
    \vspace{-0.1in}
    \caption{\textbf{Qualitative comparison among different models.} Model A performs the worst, failing to move shadows and leaving obvious artifacts. Models B and C reduce artifacts but still struggle to move shadows effectively. Model D performs the best.}
    \vspace{-0.15in}
    \label{fig:ablation}
\end{figure}

We perform ablation studies to illustrate the impact of our primary design choices. Every model is trained at a resolution of $256 \times 256$ for an equal number of iterations. As shown in \cref{tab:ablation} and \cref{fig:ablation}, our designs are proven to be effective. We begin by highlighting the importance of utilizing a pretrained image-to-video model. To this end, we train an alternative model based on a pretrained text-to-image model (Model A) using our synthetic data across three tasks. Notably, the architecture and parameters of the text-to-image model are the same as those of the image-to-video model, with the exception that the former is pre-trained on single-frame images. Model A performs significantly worse compared to Model C, which is trained using a pretrained image-to-video model with the identical dataset and tasks, as evidenced by both quantitative and qualitative evaluations. Furthermore, incorporating object removal and insertion tasks into the synthetic dataset maintains similar performance for our primary task while enabling inference over three tasks (Model B vs Model C). Finally, training with real-world video data improves performance (Model C vs Model D), particularly enhancing background occlusion filling (e.g., the second example) and the consistency of lighting and shadow adjustments.
\section{Conclusion} 
\label{sec:conclusion}

In this paper, we presented \textit{ObjectMover}, a novel framework for object movement that integrates a sequence-to-sequence modeling using a video diffusion model. Our method offers a unified, single-stage solution that addresses the challenges of object identity preservation, lighting consistency, and realistic harmonization into new environments. For task-specific model finetuning, we developed a data generation pipeline utilizing a game engine to produce high-quality data pairs. Additionally, we propose a multi-task learning approach that allows for training on real-world video data, enhancing model generalization. The trained model can also be used for object removal and insertion tasks. Extensive experimental results demonstrate that our model not only attains but surpasses the effectiveness of current leading methods.

{
    \small
    \bibliographystyle{ieeenat_fullname}
    \bibliography{main}
}

\clearpage
\centerline{\Large \textbf{Appendix}}
\renewcommand*{\thesection}{\Alph{section}}
\setcounter{section}{0}
\vspace{4pt}

\section{Details on Compared Methods}
We provide more implementation details of two comparison methods, i.e., DragAnything and 3DiT. 
\paragraph{DragAnything~\cite{drag_anything}}
DragAnything is a trajectory-based video generation model that allows specifying one or more objects, enabling the corresponding objects to move according to the trajectory in the generated video. However, while this method can control the positions of the generated objects to match the coordinates in the trajectory, it cannot control other elements, such as keeping the background stationary. To maintain the background as much as possible, we need to select a point in the background region and assign it a stationary trajectory to control the relative stability of the background. In our implementation, we strive to keep the background stationary. Specifically, we design two trajectories. The first is the foreground trajectory, which controls the movement of the target object. The start point is the original center position of the object, and the end point is the center of the target position of the object. The key points of the trajectory are obtained through linear interpolation. The second is the background trajectory, where we set the trajectories of a background point in the background region to be stationary, thereby maintaining the stability of the background. To automatically select a background point, we identify the location within the background mask that is farthest from both the foreground object and the image boundaries. This is achieved by computing the Euclidean distance of each background pixel to the nearest foreground pixel and the image borders, then selecting the point with the maximum minimum distance. 
Nonetheless, despite our effort, this can still result in detail-level jitter or control failures, as shown in the last example of~\cref{fig:supp_move_A}.

\paragraph{3DiT~\cite{3dit}}
3DiT is a text-conditioned image editing method that cannot use bounding boxes to precisely control the objects to be moved and their target positions. Instead, it requires a textual prompt to describe the objects and the coordinates of the target positions. To address this, we employ an image caption model to generate text labels for each cropped object in our evaluation set, which are then used as prompt instructions. For the target position coordinates, we use the center points of the target bounding boxes.

\section{Video Dataset Pipeline}
We use an internal video dataset as the real-world video source. Note that the dataset pipeline can be applied to any video dataset.
For processing, we utilize SAM2~\cite{sam2} to segment the videos and obtain consistent object labels across frames. Then, we filter out objects with masks that are too small and those that do not appear simultaneously in both frames. Finally, we obtain approximately 800,000 image groups, each containing two frames and the corresponding mask image for one object.

\begin{figure}[h]
    \centering
    \includegraphics[width=1\linewidth]{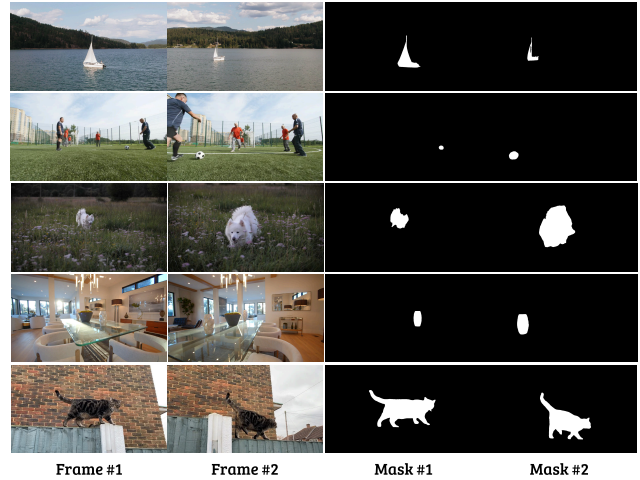}
    \caption{\textbf{Video dataset samples.} Frames and corresponding masks from our video dataset, processed with the SAM2~\cite{sam2} to ensure consistent object labeling across frames.}
    \label{fig:supp_video_data}
\end{figure}

\begin{figure}[h]
    \centering
    \includegraphics[width=1\linewidth]{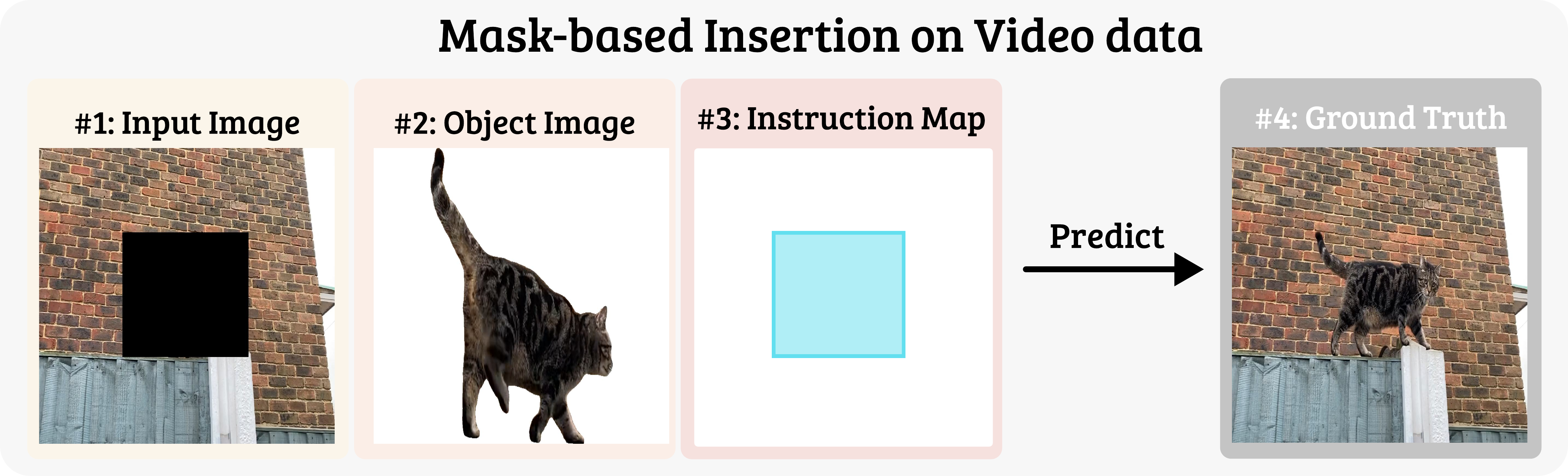}
    \caption{\textbf{Mask-based insertion on video data.} \#1: Input image with the object masked out. \#2: Isolated object image. \#3: Instruction map indicating where to place the object. \#4: Ground-truth image for prediction.}
    \label{fig:mask_based}
\end{figure}

\begin{figure}[h]
    \centering
    \includegraphics[width=1\linewidth]{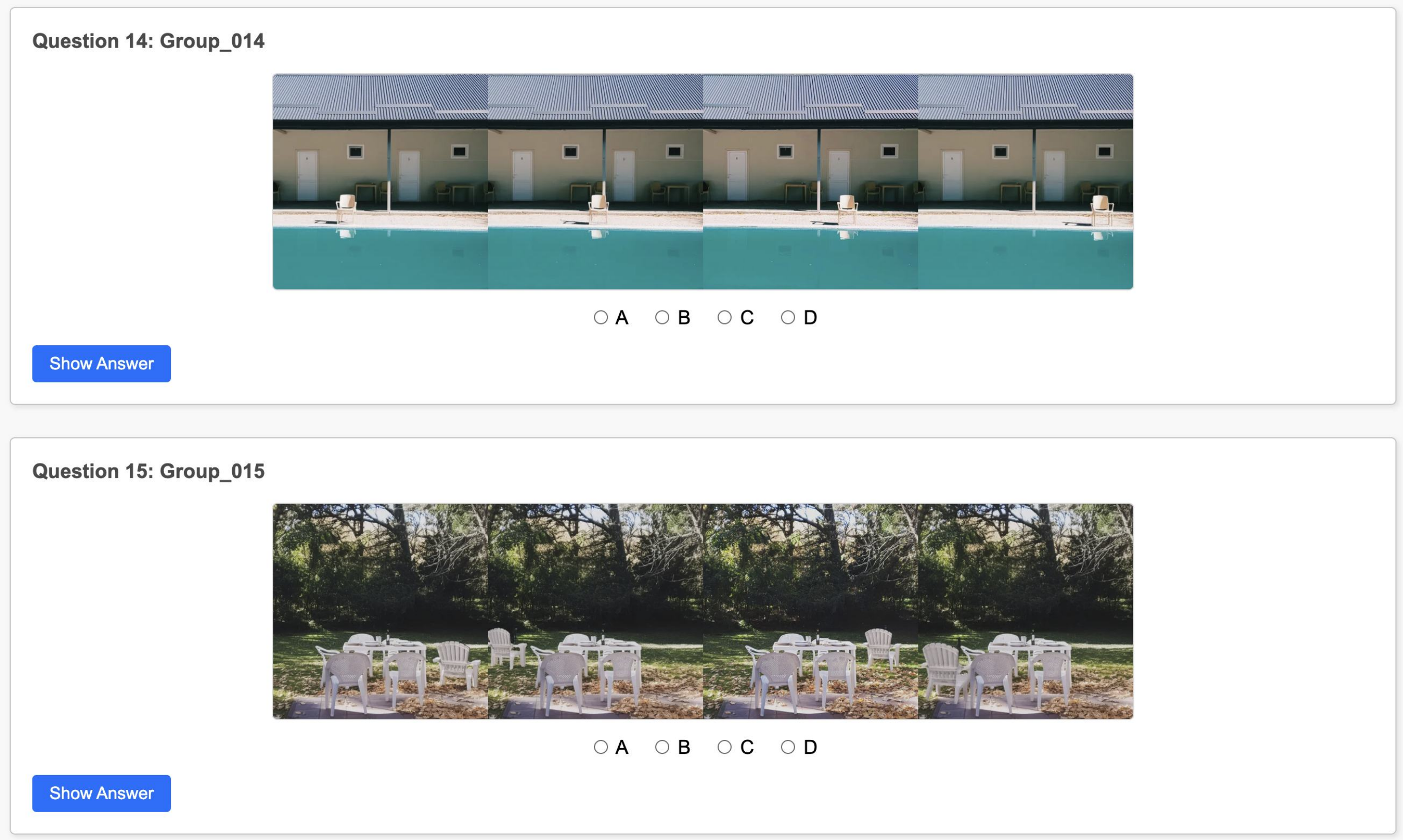}
    \caption{\textbf{Interface of the additional user study.} Samples from our ``find-the-real-image" game designed to assess the realism of synthesized images. Each participant is shown a set of images and asked to identify the original.}
    \label{fig:game}
\end{figure}

\begin{figure}
    \centering
    \includegraphics[width=1\linewidth]{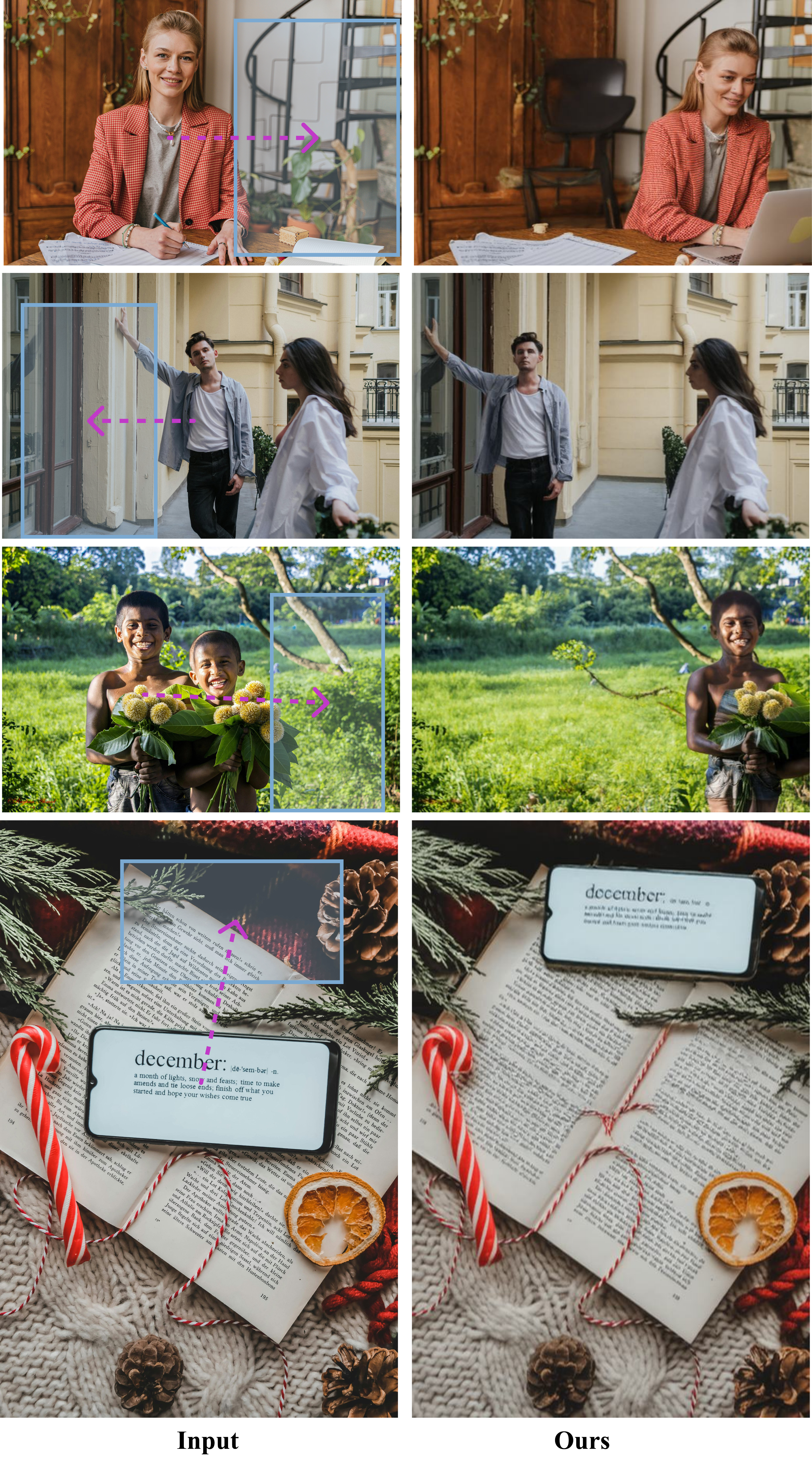}
    \caption{\textbf{Illustration of representative failure cases.} Rows 1 and 2 show unintended pose alterations when moving non-rigid objects (e.g., humans), where the generated pose significantly deviates from the original. Row 3 illustrates the disappearance of nearby objects when one object is moved closely past another. Row 4 shows text distortion after object movement, a common limitation inherent in latent diffusion models. }
    \label{fig:failure_cases}
\end{figure}

\begin{figure*}[t]
    \centering
    \includegraphics[width=0.95\linewidth]{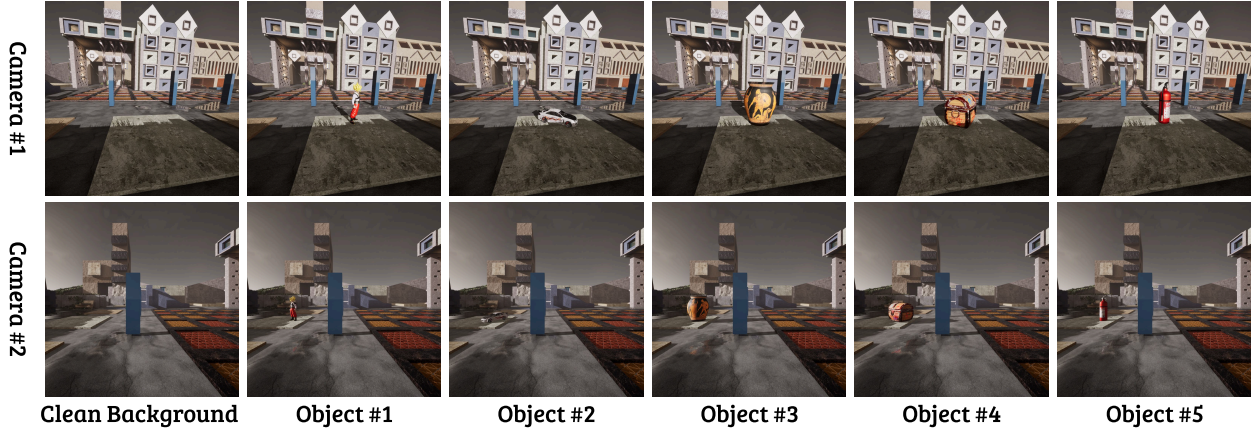}
    \caption{\textbf{Image samples of synthetic data.} Display of synthetic scenes with object placements across varying camera angles.}
    \label{fig:ue_1}
\end{figure*}

\begin{figure*}[t]
    \centering
    \includegraphics[width=0.95\linewidth]{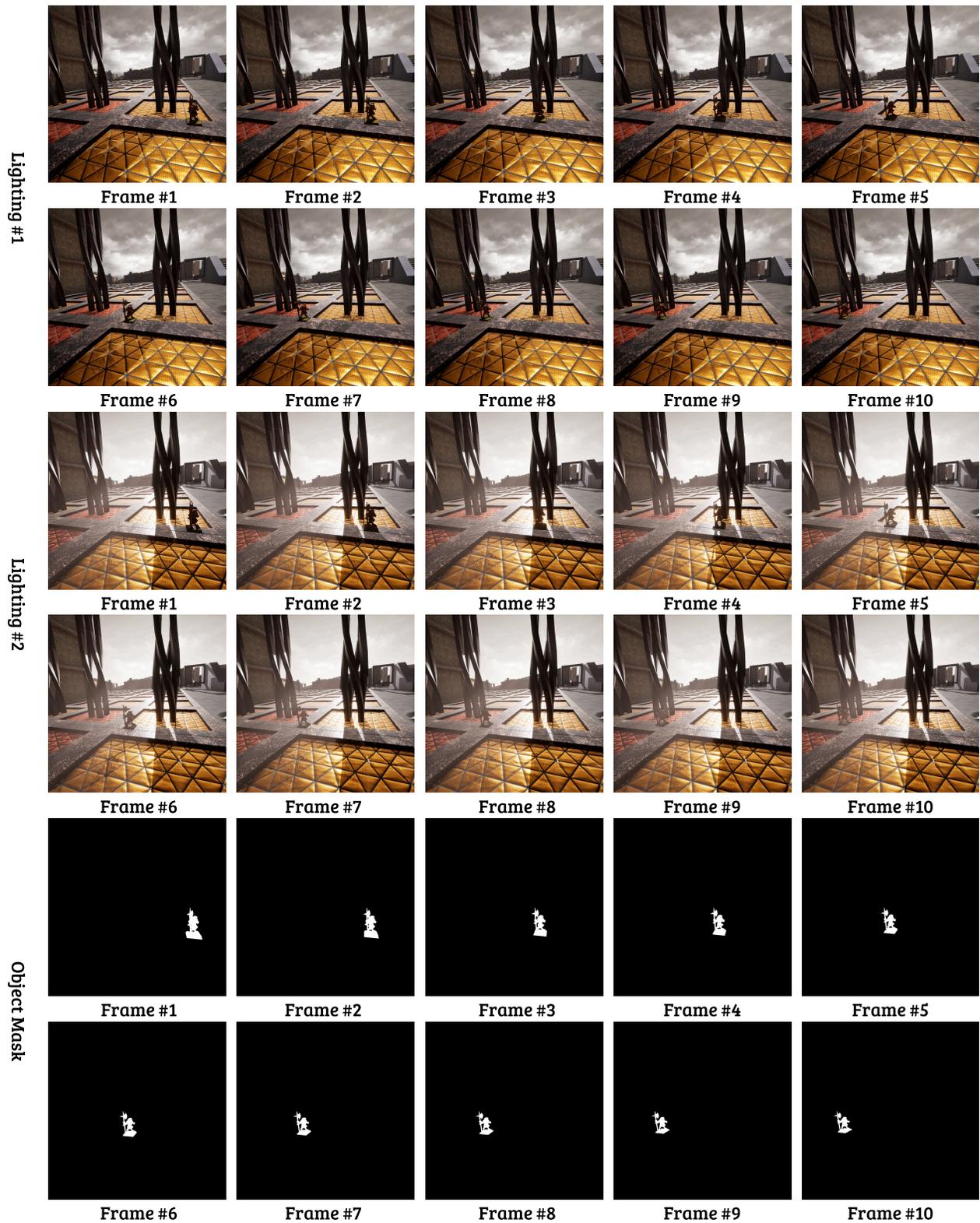}
    \caption{\textbf{Full sequence examples of synthetic data.} We show two sequences from our synthetic dataset with an object placed in different locations with two lighting conditions. The last two rows present the object masks.}
    \label{fig:ue_2}
\end{figure*}

\cref{fig:supp_video_data} shows some sample data from the video dataset. As mentioned, in the video dataset, other objects or backgrounds, except for the main subject, mostly change, making it difficult to directly use them for training movement tasks. However, we use the video data to train on a mask-based object insertion task. As demonstrated in \cref{fig:mask_based}, during the training process, the object from frame $\#1$ is extracted using mask $\#1$ and serves as the object image, coupled with a foreground-masked frame $\#2$ as the input, to predict a complete frame $\#2$.

Note that our model is trained on video and CG data, while it is evaluated on image data. Hence the training and evaluation data are completely different with no overlap or data leakage issues. \textit{ObjMove-A} is manually captured using DSLR while \textit{ObjMove-B} is web data \textbf{without ground-truth}, which theoretically prevents data leakage. 

\section{Samples of Synthetic Data}
\cref{fig:ue_1} shows some rendered images of different objects in background scenes with varying camera views. We also display their corresponding clean background images where no object is located in the region of interest. These clean background images support our mask-free object removal and insertion training. \cref{fig:ue_2} presents examples of the full sequence rendering. The first four rows illustrate sequences of the same scene and object in different positions, albeit under different lighting conditions. The last two rows display the corresponding object masks, which are directly obtained through rendering. Notably, these masks represent the amodal extent of the objects without considering occlusion relationships. This approach aids the model in learning to determine whether an object should be occluded when a mask overlaps with another object.

\section{Additional User Study: Find the Real Image}
We conduct an additional user study where users are asked to find the real image among four images, of which three are generated by our method by moving the object of interest to different locations. The results reveal that users incorrectly identify the real (input) image $70\%$ of the time, demonstrating our method’s ability to generate realistic images that effectively obscure artifacts. Samples of this game are illustrated in \cref{fig:game}, and we also provide a web link \textcolor{pink}{\href{https://xinyu-andy.github.io/ObjMover/self_evaluation_game/click_to_play_the_game.html}{here}}
to play the game interactively.

\section{More Results}
\begin{figure*}[t]
    \centering
    \includegraphics[width=0.95\linewidth]{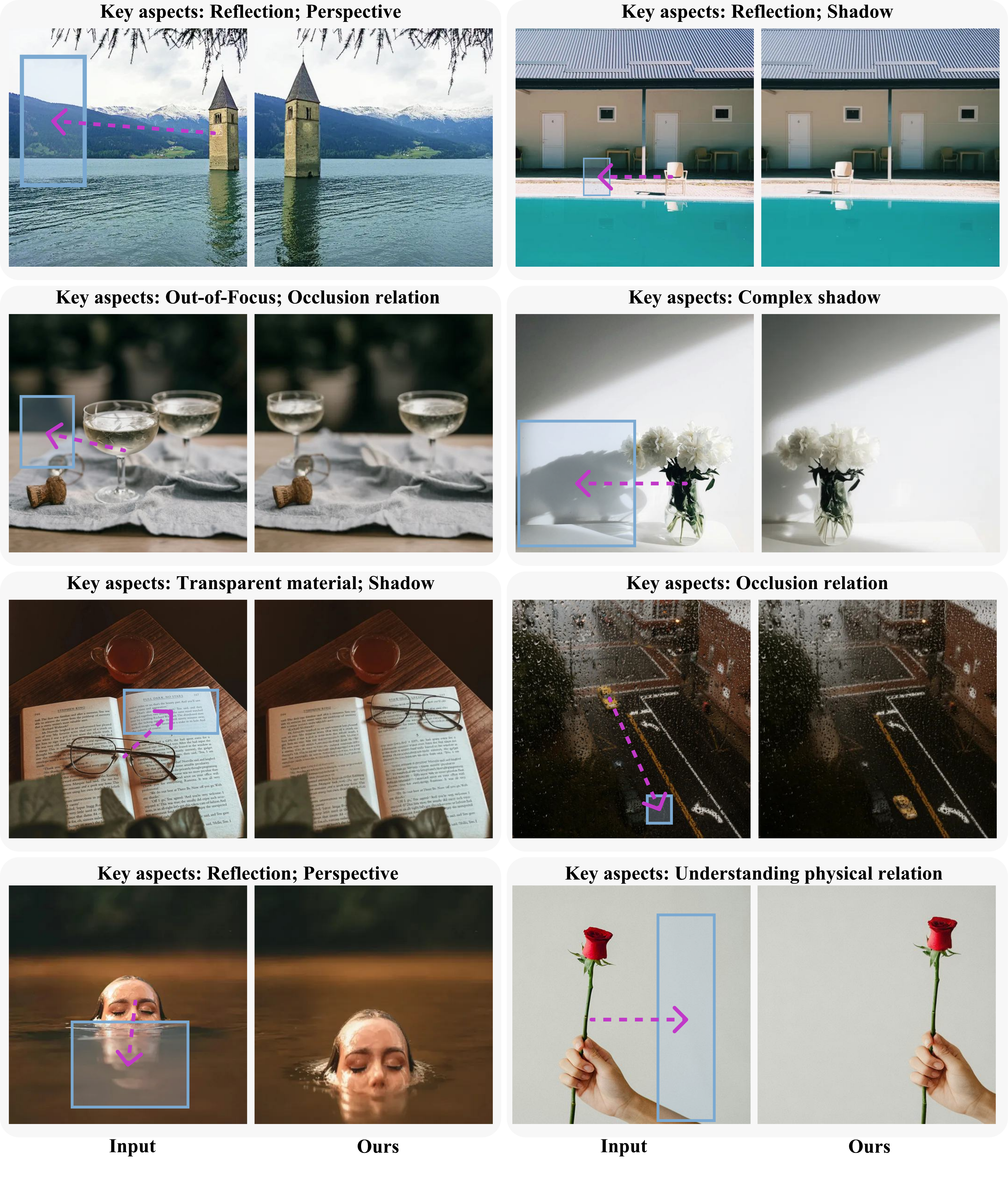}
    \caption{\textbf{Qualitative results on object movement.} Key aspects to focus on are annotated above each image to highlight the model's ability.}
    \vspace{-0.2in}
    \label{fig:supp_move}
\end{figure*}

\begin{figure*}[t]
    \centering
    \includegraphics[width=0.95\linewidth]{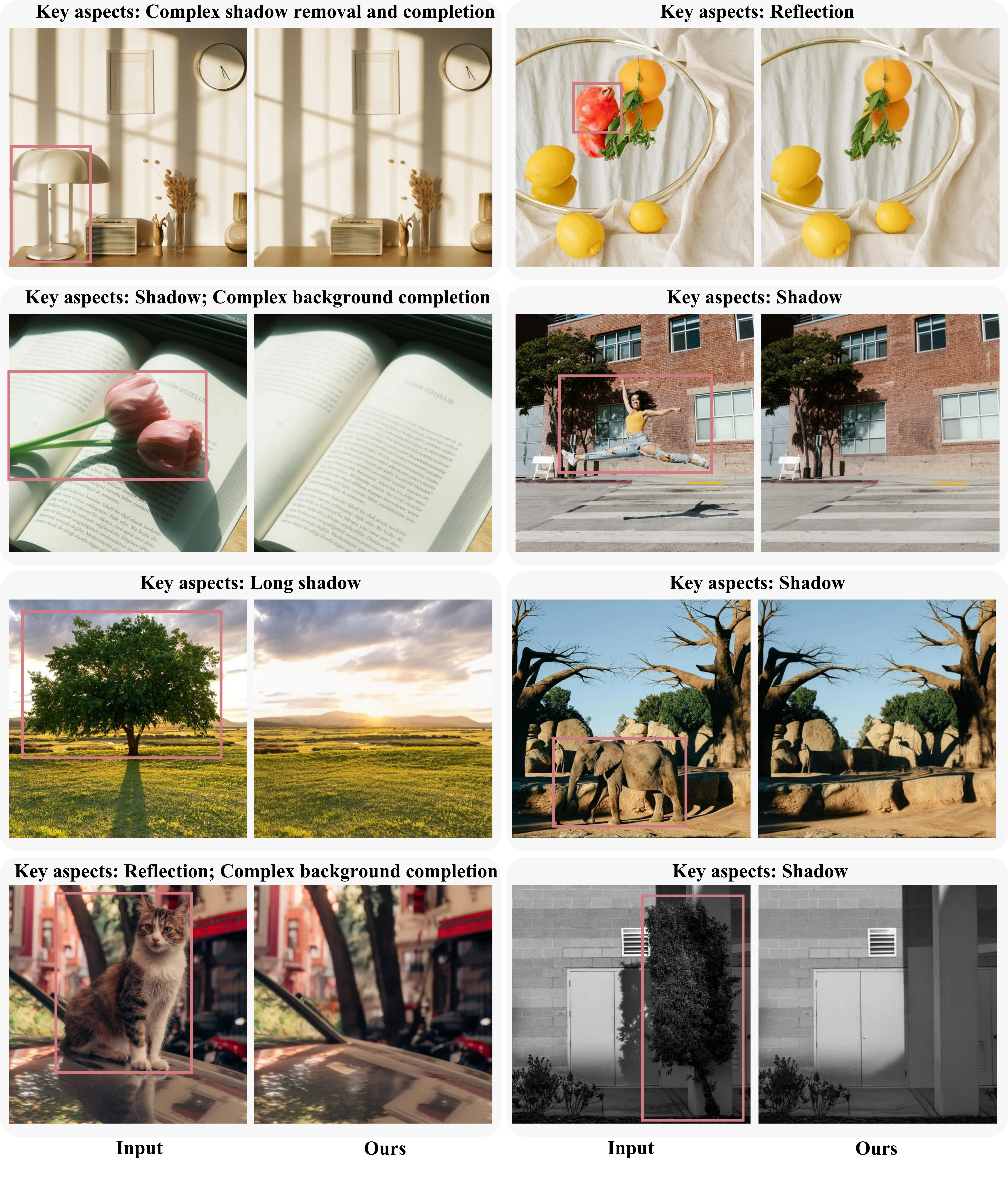}
    \caption{\textbf{Qualitative results on object removal.} Key aspects to focus on are annotated above each image to highlight the model's ability.}
    \label{fig:supp_remove}
\end{figure*}

\begin{figure*}[t]
    \centering
    \includegraphics[width=0.95\linewidth]{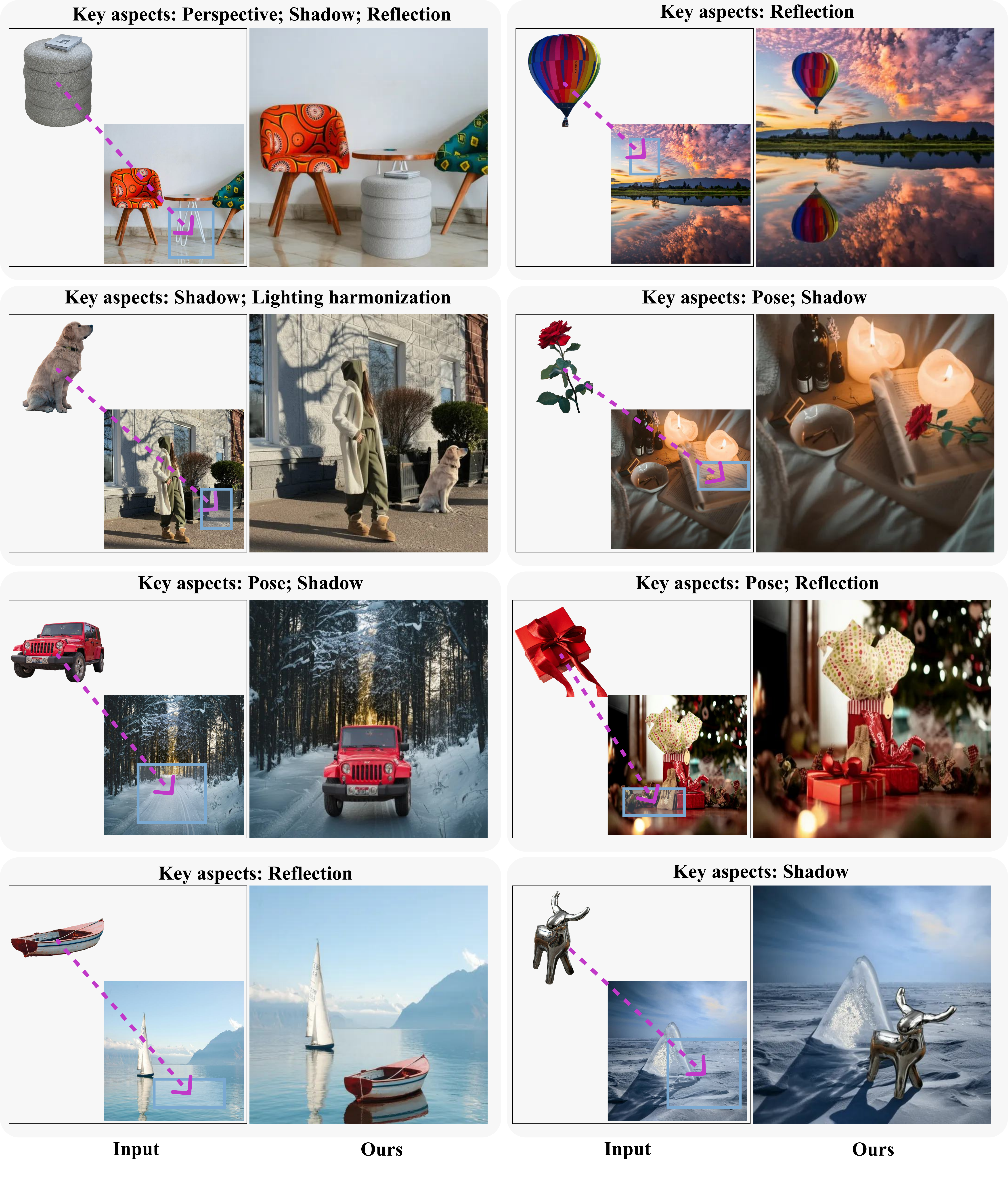}
    \caption{\textbf{Qualitative results on object insertion.} Key aspects to focus on are annotated above each image to highlight the model's ability.}
    \label{fig:supp_insert}
\end{figure*}

We provide additional qualitative results of our model on in-the-wild internet images. \cref{fig:supp_move}, \cref{fig:supp_remove}, and \cref{fig:supp_insert} respectively showcase more results of our model on object movement, removal, and insertion. For each result, we annotate key aspects above the images to better demonstrate the capabilities of our model.

\section{More Comparisons}
\begin{figure*}[t]
    \centering
    \includegraphics[width=1\linewidth]{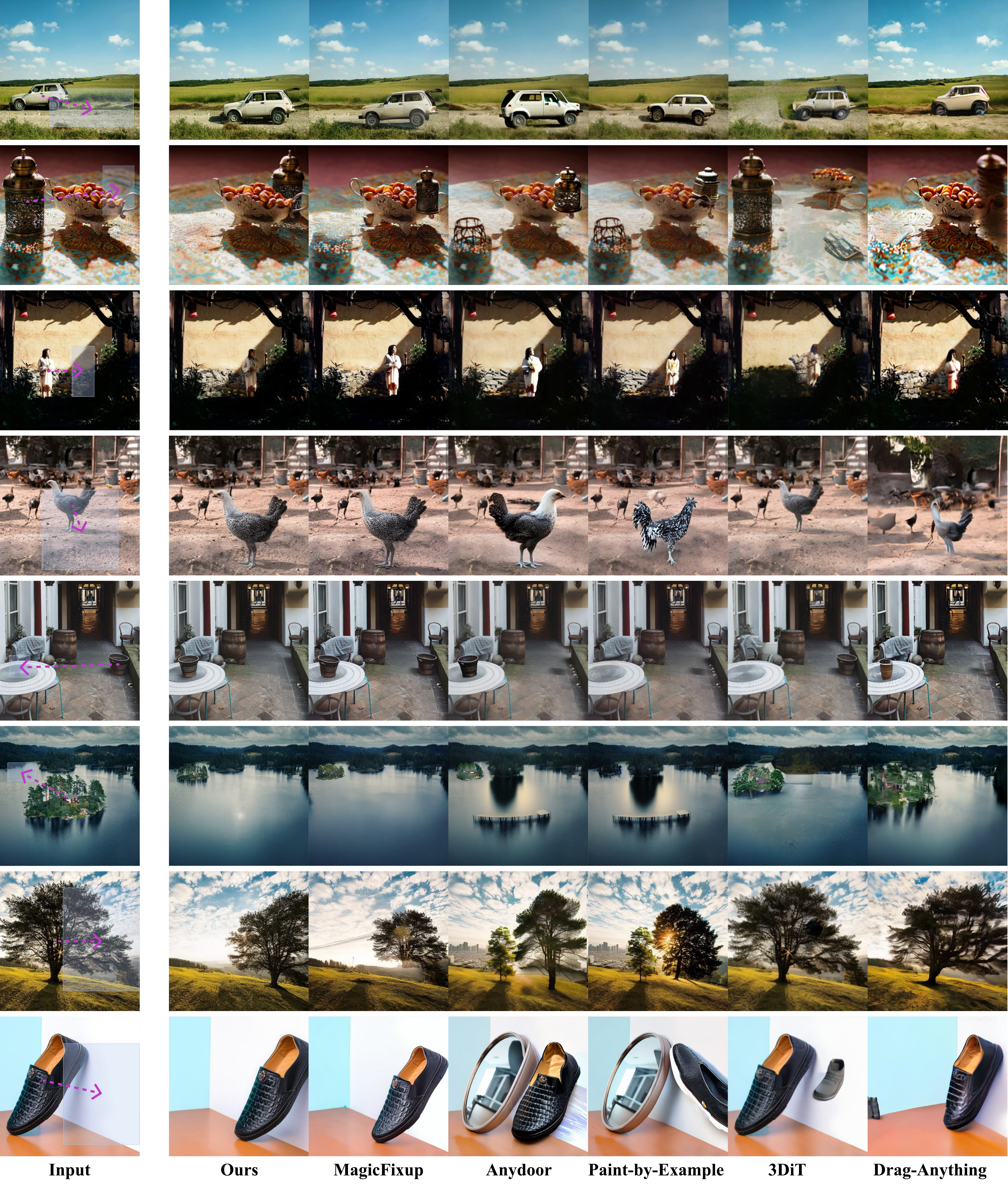}
    \caption{\textbf{Qualitative comparisons on object movement.} Our method consistently outperforms state-of-the-art methods.}
    \vspace{-0.2in}
    \label{fig:supp_move_A}
\end{figure*}

\begin{figure*}[t]
    \centering
    \includegraphics[width=1\linewidth]{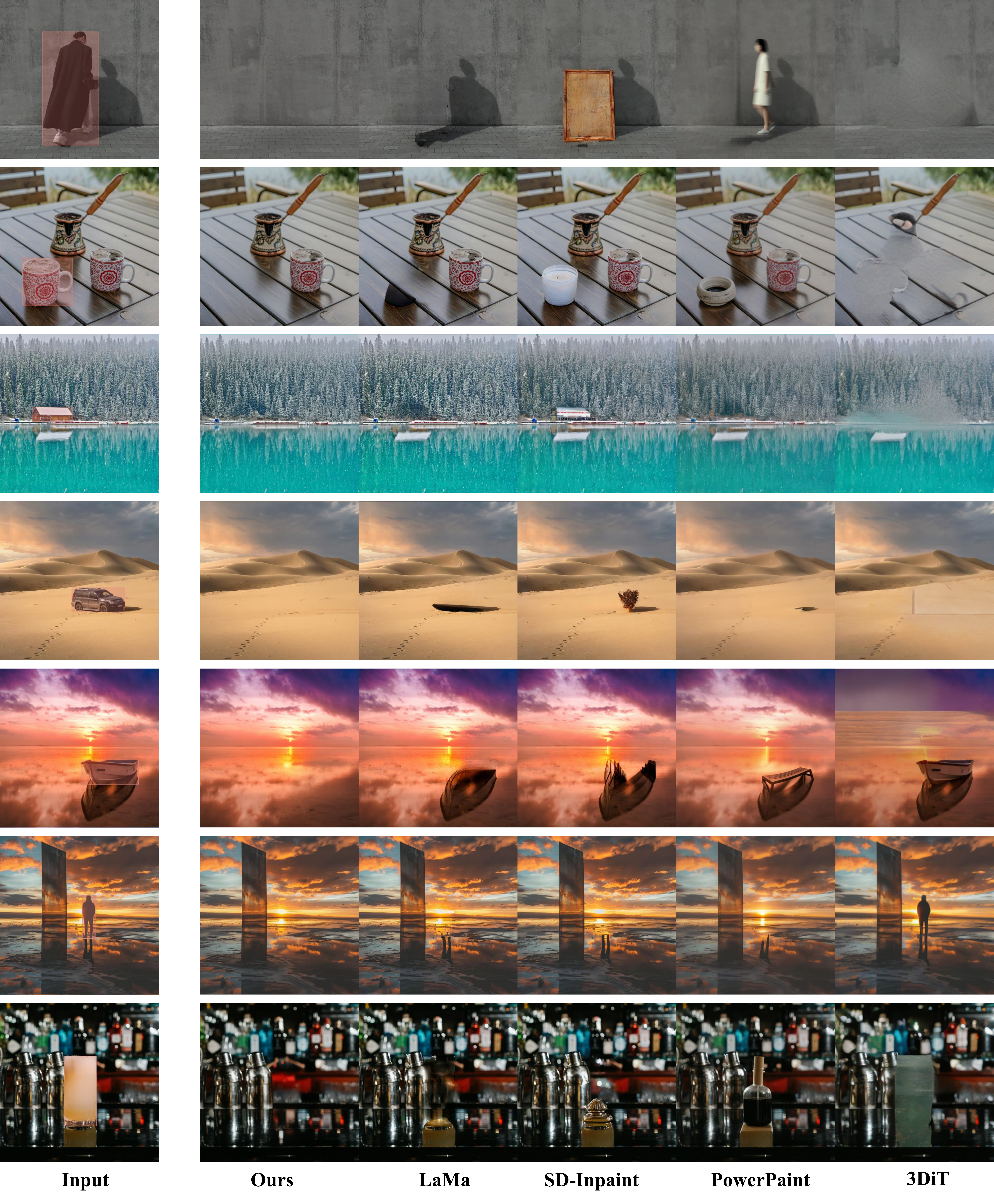}
    \caption{\textbf{Qualitative comparisons on object removal.} Our method consistently outperforms state-of-the-art methods.}
    \vspace{-0.2in}
    \label{fig:supp_remove_A}
\end{figure*}

\begin{figure*}[t]
    \centering
    \includegraphics[width=1\linewidth]{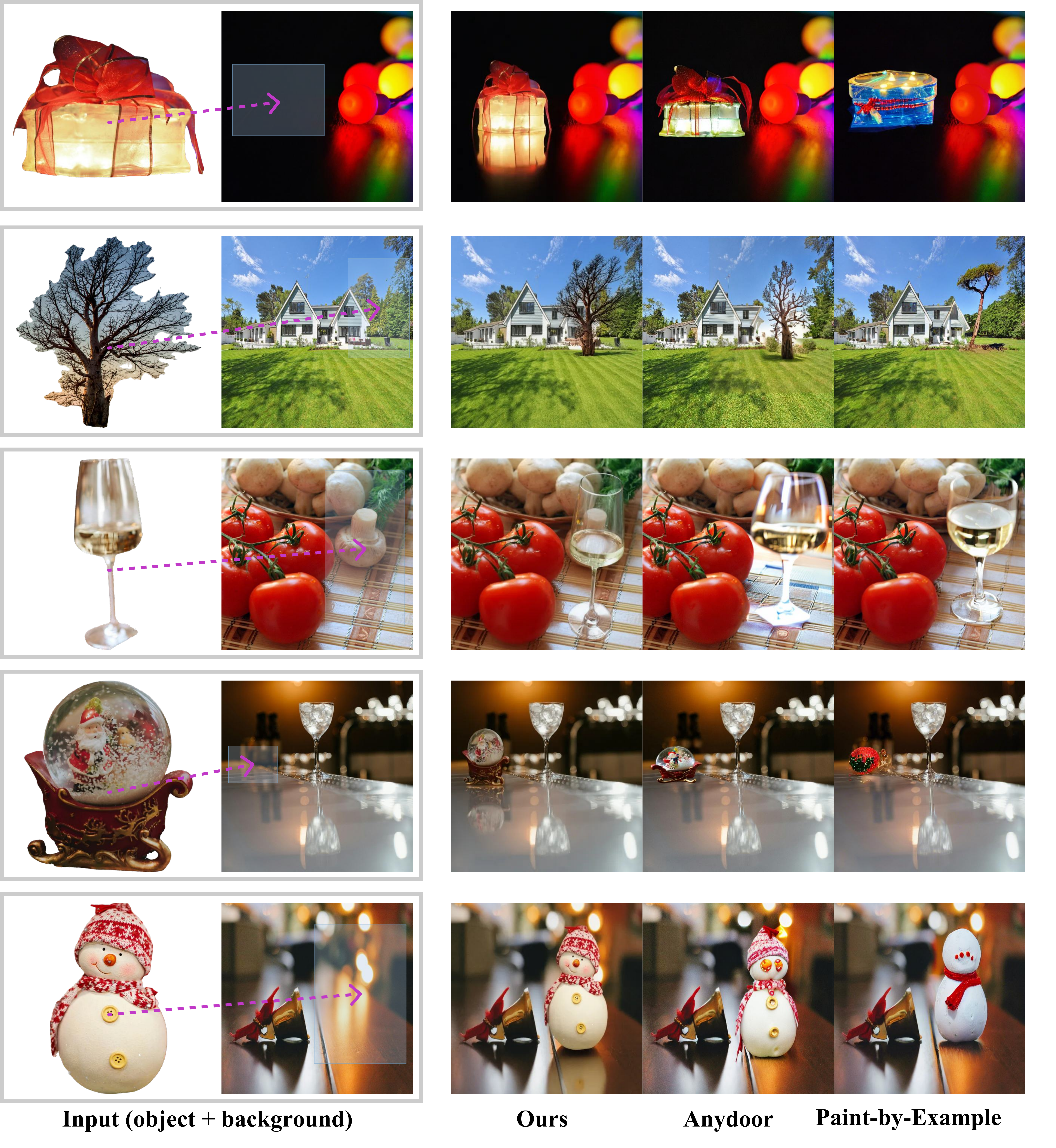}
    \caption{\textbf{Qualitative comparisons on object insertion.} Our method consistently outperforms state-of-the-art methods.}
    \vspace{-0.2in}
    \label{fig:supp_insert_A}
\end{figure*}

\begin{figure*}[t]
    \centering
    \includegraphics[width=1\linewidth]{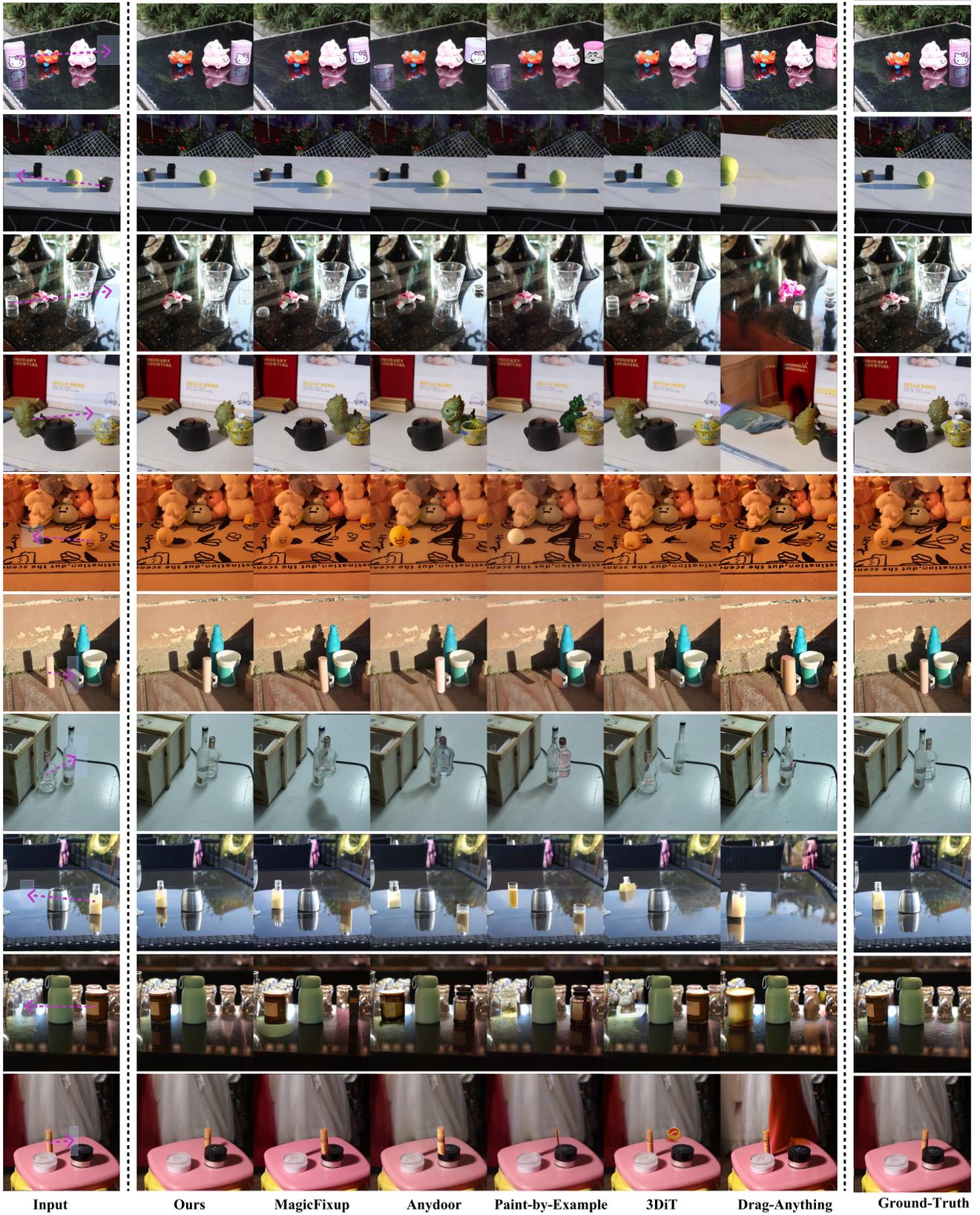}
    \caption{\textbf{Qualitative comparisons on object movement.} Our method consistently outperforms state-of-the-art methods.}
    \vspace{-0.2in}
    \label{fig:supp_move_B}
\end{figure*}

\begin{figure*}[t]
    \centering
    \includegraphics[width=1\linewidth]{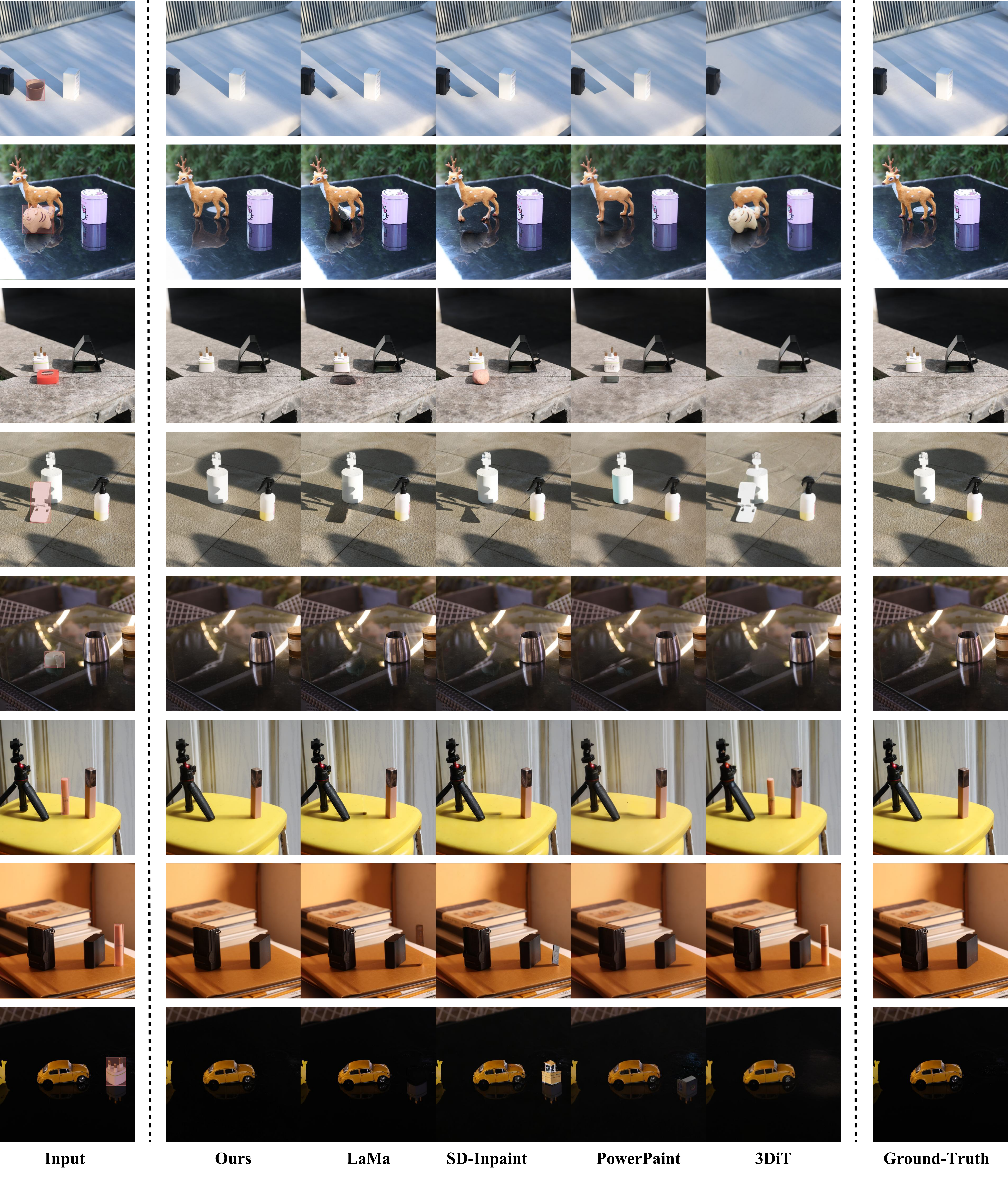}
    \caption{\textbf{Qualitative comparisons on object removal.} Our method consistently outperforms state-of-the-art methods.}
    \vspace{-0.2in}
    \label{fig:supp_remove_B}
\end{figure*}

\begin{figure*}[t]
    \centering
    \includegraphics[width=1\linewidth]{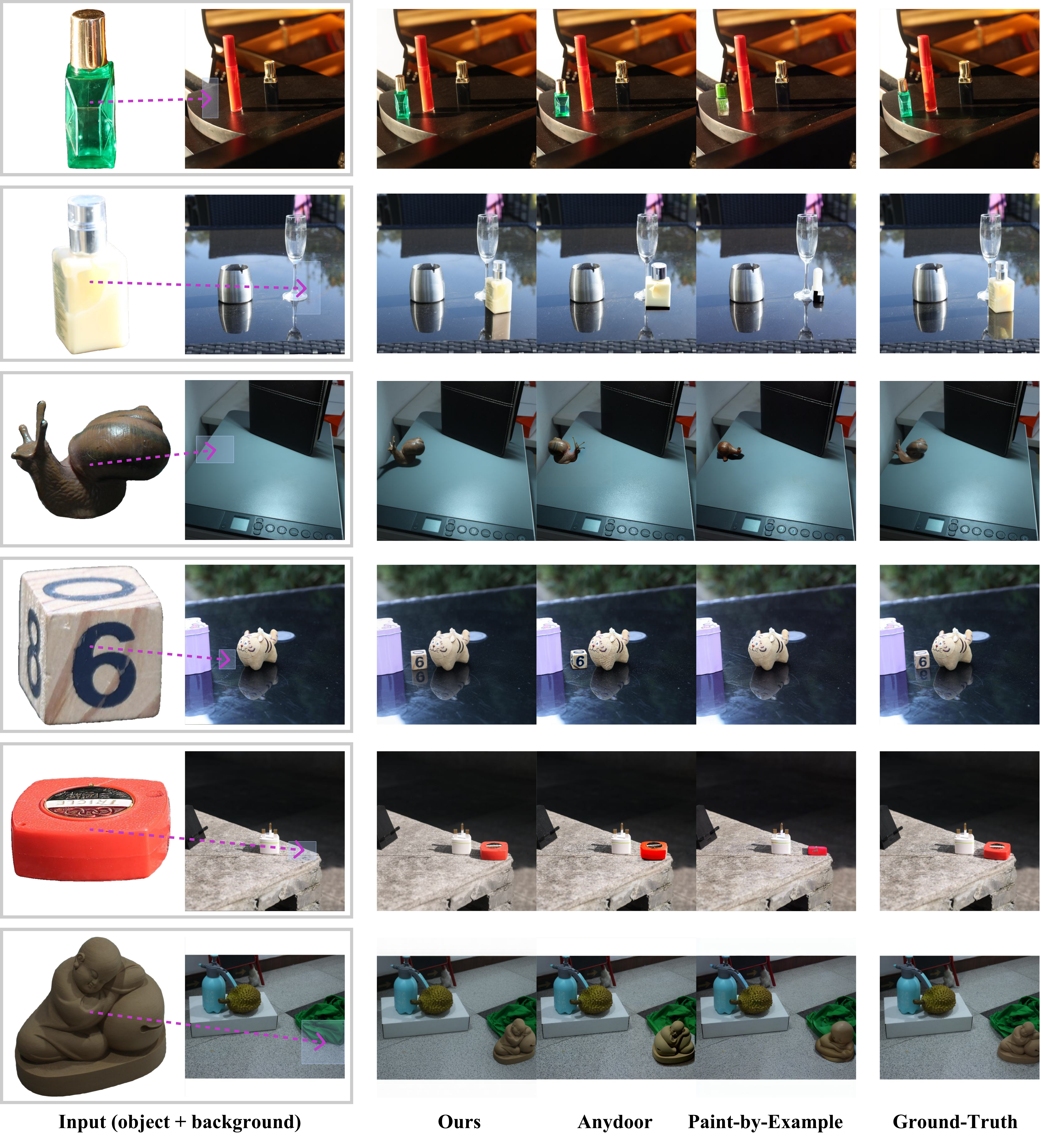}
    \caption{\textbf{Qualitative comparisons on object insertion.} Our method consistently outperforms state-of-the-art methods.}
    \vspace{-0.2in}
    \label{fig:supp_insert_B}
\end{figure*}

We present additional comparison results between our method and other approaches. \cref{fig:supp_move_A} and \cref{fig:supp_remove_A} respectively show the comparison results for the movement and removal tasks on \textit{ObjMove-B}. \cref{fig:supp_insert_A} displays the insertion results on in-the-wild image pairs. Additionally, \cref{fig:supp_move_B}, \cref{fig:supp_remove_B}, and \cref{fig:supp_insert_B} illustrate the movement, removal, and insertion results on \textit{ObjMove-A}, where a reference ground-truth image is also provided.

\section{Limitations and Future Work}
While our method achieves excellent results across three tasks---object movement, removal, and insertion---it still possesses certain limitations. Figure~\ref{fig:failure_cases} illustrates several failure cases, categorized into three main scenarios:

\begin{enumerate}
    \item \textbf{Unintended Pose Alterations.} 
    Our design philosophy emphasizes maintaining the object's original pose as consistently as possible, only automatically adjusting the pose when necessary for harmonious integration into the new environment. This strategy generally ensures stable and robust performance. However, for non-rigid objects (e.g., humans), generated results sometimes exhibit significant and unintended pose alterations, occasionally introducing new content (rows 1 and 2 in Figure~\ref{fig:failure_cases}). We suspect this primarily arises from the abundance of human-motion examples in real video datasets, which bias the model towards pose variability. To address this, we plan to incorporate meta-information regarding relative object-camera poses into our synthetic dataset and conditionally train the model based on this information. This enhancement will enable explicit 3D control, allowing precise, user-directed pose manipulation.
    
    \item \textbf{Disappearance of Nearby Objects.} 
    When an object is moved closely past another object (row 3 in Figure~\ref{fig:failure_cases}), the nearby object occasionally disappears. We attribute this to a lack of examples where one object explicitly crosses over another within our synthetic training data. This limitation can easily be resolved by augmenting the dataset with relevant scenarios.

    \item \textbf{Text Distortion.} 
    For objects containing text (row 4 in Figure~\ref{fig:failure_cases}), moving the object often results in distorted text. This is a common limitation in latent diffusion models caused by insufficient reconstruction capabilities of the VAE.
\end{enumerate}

Moreover, our method exhibits relatively slow inference speed. On a single NVIDIA A100 GPU, inferring an image with a resolution of 512$\times$512 requires approximately 20 seconds, which is slower than other U-Net-based approaches. However, in future work, we aim to reduce the inference cost by employing model distillation and diffusion distillation techniques~\cite{dmd,improved_dmd,consistency_model}, thereby enhancing the practical applicability of our approach in real-world scenarios.

\end{document}